\documentclass[review]{elsarticle}
\usepackage[letterpaper, left=2cm, right=2cm, top=2cm, bottom=2cm]{geometry}
\usepackage{hyperref}
\usepackage{url}
\usepackage{breakurl}
\usepackage{graphicx,color}
\usepackage{paralist}
\usepackage{amsmath,amsfonts,amssymb}
\usepackage{subfigure}
\usepackage{algorithm2e}
\usepackage{comment}
\usepackage{multirow}
\usepackage[square]{natbib}

\newcommand{\bx}{{\bf x}}
\newcommand{\bc}{{\bf c}}
\newcommand{\bv}{{\bf v}}

\begin{document}

\title{Improving the Performance of K-Means for Color Quantization}
\author{%
        M. Emre Celebi\\
        Department of Computer Science\\Louisiana State University, Shreveport, LA, USA\\
        \href{mailto:ecelebi@lsus.edu}{ecelebi@lsus.edu}
       }

\begin{abstract}
Color quantization is an important operation with many applications in graphics and image processing. Most quantization methods are essentially based on data clustering algorithms. However, despite its popularity as a general purpose clustering algorithm, k-means has not received much respect in the color quantization literature because of its high computational requirements and sensitivity to initialization. In this paper, we investigate the performance of k-means as a color quantizer. We implement fast and exact variants of k-means with several initialization schemes and then compare the resulting quantizers to some of the most popular quantizers in the literature. Experiments on a diverse set of images demonstrate that an efficient implementation of k-means with an appropriate initialization strategy can in fact serve as a very effective color quantizer.
\end{abstract}

\maketitle

\section{Introduction}
\label{sec_intro}

True-color images typically contain thousands of colors, which makes their display, storage, transmission, and processing problematic. For this reason, color quantization (reduction) is commonly used as a preprocessing step for various graphics and image processing tasks. In the past, color quantization was a necessity due to the limitations of the display hardware, which could not handle over 16 million possible colors in 24-bit images. Although 24-bit display hardware has become more common, color quantization still maintains its practical value \citep{Brun02}. Modern applications of color quantization in graphics and image processing include:
\begin{inparaenum}[(i)]
 \item compression \citep{Yang98},
 \item segmentation \citep{Deng01a},
 \item text localization/detection \citep{Sherkat05},
 \item color-texture analysis \citep{Sertel09},
 \item watermarking \citep{Kuo07},
 \item non-photorealistic rendering \cite{Wang10}, and
 \item content-based retrieval \citep{Deng01b}.
\end{inparaenum}
\par
The process of color quantization is mainly comprised of two phases: palette design (the selection of a small set of colors that represents the original image colors) and pixel mapping (the assignment of each input pixel to one of the palette colors). The primary objective is to reduce the number of unique colors, $N'$, in an image to $K$ ($K \ll N'$) with minimal distortion. In most applications, 24-bit pixels in the original image are reduced to 8 bits or fewer. Since natural images often contain a large number of colors, faithful representation of these images with a limited size palette is a difficult problem.
\par
Color quantization methods can be broadly classified into two categories \citep{Xiang07}: image-independent methods that determine a universal (fixed) palette without regard to any specific image \citep{Gentile90,Mojsilovic01}, and image-dependent methods that determine a custom (adaptive) palette based on the color distribution of the images. Despite being very fast, image-independent methods usually give poor results since they do not take into account the image contents. Therefore, most of the studies in the literature consider only image-dependent methods, which strive to achieve a better balance between computational efficiency and visual quality of the quantization output.
\par
Numerous image-dependent color quantization methods have been developed in the past three decades. These can be categorized into two  families: preclustering methods and postclustering methods \citep{Brun02}. Preclustering methods are mostly based on the statistical analysis of the color distribution of the images. Divisive preclustering methods start with a single cluster that contains all $N$ image pixels. This initial cluster is recursively subdivided until $K$ clusters are obtained. Well-known divisive methods include median-cut \citep{Heckbert82}, octree \citep{Gervautz88}, variance-based method \citep{Wan90}, binary splitting \citep{Orchard91}, greedy orthogonal bipartitioning \citep{Wu91}, optimal principal multilevel quantizer \citep{Wu92}, center-cut \citep{Joy93}, and rwm-cut \citep{Yang96}. More recent methods can be found in \citep{Hsieh00,Cheng01,Lo03,Sirisathitkul04,Kanjanawanishkul05}. On the other hand, agglomerative preclustering methods \citep{Equitz89,Balasubramanian91,Xiang94,Velho97,Brun00,Franti06} start with $N$ singleton clusters each of which contains one image pixel. These clusters are repeatedly merged until $K$ clusters remain. In contrast to preclustering methods that compute the palette only once, postclutering methods first determine an initial palette and then improve it iteratively. Essentially, any data clustering method can be used for this purpose. Since these methods involve iterative or stochastic optimization, they can obtain higher quality results when compared to preclustering methods at the expense of increased computational time. Clustering algorithms adapted to color quantization include k-means \citep{Kasuga00,Huang04,Hu07,Hu08a}, minmax \citep{Xiang97}, competitive learning \citep{Uchiyama94a,Verevka95,Scheunders97,Celebi09,Celebi10b}, fuzzy c-means \citep{Ozdemir02,Schaefer09}, BIRCH \citep{Bing04}, and  self-organizing maps \citep{Dekker94,Papamarkos02,Chang05}.
\par
In this paper, we investigate the performance of the k-means (KM) clustering algorithm \citep{Lloyd82} as a color quantizer. We implement several efficient KM variants each one with a different initialization scheme and then compare these to some of the most popular color quantizers on a diverse set of images. The rest of the paper is organized as follows. Section \ref{sec_kmeans_quant} describes the conventional KM algorithm, a novel way to accelerate it, and several generic schemes to initialize it. Section \ref{sec_exp} describes the experimental setup, demonstrates the computational advantage of the accelerated KM algorithm over the conventional one, and compares the accelerated KM variants with various initialization schemes to other color quantization methods. Finally, Section \ref{sec_conc} gives the conclusions.

\section{Color Quantization Using K-Means Clustering Algorithm}
\label{sec_kmeans_quant}

\subsection{K-Means Clustering Algorithm}
\label{sec_kmeans}

The KM algorithm is inarguably one of the most widely used methods for data clustering \citep{Gan07}. Given a data set $X = \left\{ \bx_1, \bx_2, \ldots, \bx_N \right\} \in \mathbb{R}^D $, the objective of KM is to partition $X$ into $K$ exhaustive and mutually exclusive clusters $S = \left\{ S_1, S_2, \ldots, S_K \right\},\;\; \bigcup\nolimits_{k = 1}^K {S_k = X},\;\; S_i \cap S_j = \emptyset$ for $1 \leq i \neq j \leq K$ by minimizing the sum of squared error (SSE):
\begin{equation}
\label{eq_sse}
 \mbox{SSE} = \sum\limits_{k = 1}^K {\sum\limits_{\bx_i \in S_k} {\left\| {\bx_i - \bc_k } \right\|_2^2 } }
\end{equation}
\noindent where $\| . \|_2$ denotes the Euclidean ($L_2$) norm and $\bc_k$ is the center of cluster $S_k$ calculated as the mean of the points that belong to this cluster. This problem is known to be NP-hard even for $K = 2$ \citep{Aloise09} or $D = 2$ \citep{Mahajan10}, but a heuristic method developed by Lloyd \citep{Lloyd82} offers a simple solution. Lloyd's algorithm starts with $K$ arbitrary centers, typically chosen uniformly at random from the data points \citep{Forgy65}. Each point is then assigned to the nearest center, and each center is recalculated as the mean of all points assigned to it. These two steps are repeated until a predefined termination criterion is met. The pseudocode for this procedure is given in Algo.\ (\ref{algo_kmeans}) (\textbf{bold} symbols denote vectors). Here, $m[i]$ denotes the membership of point $\bx_i$, i.e.\ index of the cluster center that is nearest to $\bx_i$.

\begin{algorithm}
\linespread{1}
\normalsize
{
\label{algo_kmeans}
\SetKwInOut{Input}{input}
\SetKwInOut{Output}{output}
\Input { $ X = \left\{ \bx_1, \bx_2, \ldots, \bx_N \right\} \in \mathbb{R}^D $ ($N \times D$ input data set) }
\Output { $ C = \left\{ \bc_1, \bc_2, \ldots, \bc_K \right\} \in \mathbb{R}^D $ ($K$ cluster centers)}
Select a random subset $C$ of $X$ as the initial set of cluster centers\;
\While { termination criterion is not met }
{
 \For{$( i = 1; i \leq N; i = i + 1 )$}
  {
   \CommentSty{Assign $\bx_i$ to the nearest cluster}\;
   $m[i] = \underset{k \in \left\{ 1, 2, \ldots, K \right\} }{\operatorname{argmin}} \left\| \bx_i - \bc_k \right\|^2$\;
  }
 \CommentSty{Recalculate the cluster centers}\;
 \For{$( k = 1; k \leq K; k = k + 1 )$}
  {
   \CommentSty{Cluster $S_k$ contains the set of points $\bx_i$ that are nearest to the center $\bc_k$}\;
   $S_k  = \left\{ {\bx_i \left| {m[i] = k} \right.} \right\}$\;
   \CommentSty{Calculate the new center $\bc_k$ as the mean of the points that belong to $S_k$}\;
   $\bc_k  = \frac{1} {{\left| {S_k } \right|}}\sum\limits_{\bx_i \in S_k } {\bx_i }$\;
  }
}
\caption{Conventional K-Means Algorithm}
}
\end{algorithm}

The complexity of KM is $\mathcal{O}(NK)$ per iteration for a fixed $D$ value. For example, in color quantization applications $D = 3$ since the clustering procedure is often performed in three-dimensional color spaces such as RGB or CIEL*a*b* \citep{Celebi10a}.
\par
From a clustering perspective KM has the following advantages:
\begin{itemize}
\renewcommand{\labelitemi}{$\diamond$}

 \item It is conceptually simple, versatile, and easy to implement.
 \item It has a time complexity that is linear in $N$ and $K$. Furthermore, numerous acceleration techniques are available in the literature \citep{Phillips02,Kanungo02,Lai08,Har-Peled04,Feldman07,Elkan03}.
 \item It is guaranteed to terminate \citep{Selim84} with a quadratic convergence rate \citep{Bottou95}.
\end{itemize}

The main disadvantages of KM are the facts that it often terminates at a local minimum \citep{Selim84} and that its output is sensitive to the initial choice of the cluster centers. From a color quantization perspective, KM has two additional drawbacks. First, despite its linear time complexity, the iterative nature of the algorithm renders the palette generation phase computationally expensive. Second, the pixel mapping phase is inefficient, since for each input pixel a full search of the palette is required to determine the nearest color. In contrast, preclustering methods often manipulate and store the palette in a special data structure (binary trees are commonly used), which allows for fast nearest neighbor search during the mapping phase. Note that these drawbacks are shared by the majority of postclustering methods and will be addressed in the following subsections.

\subsection{Accelerating the K-Means Algorithm}
\label{sec_wsmeans}

In order to make it more suitable for color quantization, we propose the following modifications to the conventional KM algorithm:

\begin{enumerate}
 \item \textbf{Data sampling}: A straightforward way to speed up KM is to reduce the amount of data, which can be achieved by subsampling the input image data. In this study, two deterministic subsampling methods are utilized. The first method involves a 2:1 subsampling in the horizontal and vertical directions, so that only 1/4-th of the input image pixels are taken into account \citep{Goldberg91}. This kind of moderate sampling has been found to be effective in reducing the computational time without degrading the quality of quantization \citep{Goldberg91, Fletcher91, Balasubramanian94, Kanjanawanishkul05}. The second method involves sampling only the pixels with unique colors. These pixels can be determined efficiently using a hash table that uses chaining for collision resolution and a universal hash function of the form: $h_{\bf a}(\bx) = \left( {\sum\nolimits_{i = 1}^3 {a_i x_i}} \right)\bmod m$, where $\bx = (x_1, x_2, x_3)$ denotes a pixel with red ($x_1$), green ($x_2$), and blue ($x_3$) components, $m$ is a prime number, and the elements of sequence ${\bf a} = (a_1, a_2, a_3)$ are chosen randomly from the set $\left\{ 0, 1, \ldots, m - 1\right\}$. This second subsampling method further reduces the image data since most images contain a large number of duplicate colors (see \S \ref{sec_criteria}).

 \item \textbf{Sample weighting}: An important disadvantage of the second subsampling method described above is that it disregards the color distribution of the original image. In order to address this problem, each point is assigned a weight that is proportional to its frequency. Note that this weighting procedure essentially generates a one-dimensional color histogram. The weights are then normalized by the number of pixels in the image to avoid numerical instabilities in the calculations. In addition, Algo.\ (\ref{algo_kmeans}) is modified to incorporate the weights in the clustering procedure.

 \item \textbf{Sort-Means algorithm}: The assignment phase of KM involves many redundant distance calculations. In particular, for each point, the distances to each of the $K$ cluster centers are calculated. Consider a point $\bx_i$, two cluster centers $\bc_a$ and $\bc_b$ and a distance metric $d$, using the triangle inequality, we have $ d(\bc_a,\bc_b) \leq d(\bx_i,\bc_a) + d(\bx_i,\bc_b) $. Therefore, if we know that $ 2d(\bx_i,\bc_a) \leq d(\bc_a,\bc_b) $, we can conclude that $ d(\bx_i,\bc_a) \leq d(\bx_i,\bc_b) $ without having to calculate $ d(\bx_i,\bc_b) $. The compare-means algorithm \citep{Phillips02} precalculates the pairwise distances between cluster centers at the beginning of each iteration. When searching for the nearest cluster center for each point, the algorithm often avoids a large number of distance calculations with the help of the triangle inequality test. The sort-means (SM) algorithm \citep{Phillips02} further reduces the number of distance calculations by sorting the distance values associated with each cluster center in ascending order. At each iteration, point $\bx_i$ is compared against the cluster centers in increasing order of distance from the center $\bc_k$ that $\bx_i$ was assigned to in the previous iteration. If a center that is far enough from $\bc_k$ is reached, all of the remaining centers can be skipped and the procedure continues with the next point. In this way, SM avoids the overhead of going through all of the centers. It should be noted that more elaborate approaches to accelerate KM have been proposed in the literature. These include algorithms based on kd-trees \citep{Kanungo02,Lai08}, coresets \citep{Har-Peled04,Feldman07}, and more sophisticated uses of the triangle inequality \citep{Elkan03}. Some of these algorithms \citep{Har-Peled04,Feldman07,Elkan03} are not suitable for low dimensional data sets such as color image data since they incur significant overhead to create and update auxiliary data structures \citep{Elkan03}. Others \citep{Kanungo02,Lai08} provide computational gains comparable to SM at the expense of significant conceptual and implementation complexity. In contrast, SM is conceptually simple, easy to implement, and incurs very small overhead, which makes it an ideal candidate for color clustering.
\end{enumerate}

We refer to the KM algorithm with the abovementioned modifications as the 'Weighted Sort-Means' (WSM) algorithm. The pseudocode for WSM is given in Algo.\ (\ref{algo_wsmeans}). Let $\gamma$ be the average over all points $p$ of the number of centers that are no more than two times as far as $p$ is from the center $p$ was assigned to in the previous iteration. The complexity of WSM is $\mathcal{O}(K^2 + K^2 \log{K} + N' \gamma)$ per iteration for a fixed $D$ value, where the terms (from left to right) represent the cost of calculating the pairwise distances between the cluster centers, the cost of sorting the centers, and the cost of comparisons, respectively. Here, the last term dominates the computational time, since in color quantization applications $K$ is a small number and furthermore $K \ll N'$. Therefore, it can be concluded that WSM is linear in $N'$, the number of unique colors in the original image. The influence of $K$ on the complexity of WSM will be empirically demonstrated in the next section. It should be noted that, when initialized with the same centers, WSM gives identical results to KM.

\begin{algorithm}
\linespread{1}
\normalsize
{
\label{algo_wsmeans}
\SetKwInOut{Input}{input}
\SetKwInOut{Output}{output}
\Input { $ X = \left\{ \bx_1, \bx_2, \ldots, \bx_{N'} \right\} \in \mathbb{R}^D $ ($N' \times D$ input data set)\\
         $ W = \left\{ w_1, w_2, \ldots, w_{N'} \right\} \in [0,1] $ ($N'$ point weights) }
\Output { $ C = \left\{ \bc_1, \bc_2, \ldots, \bc_K \right\} \in \mathbb{R}^D $ ($K$ cluster centers) }
Select a random subset $C$ of $X$ as the initial set of cluster centers\;
\While { termination criterion is not met }
{
 \CommentSty{Calculate the pairwise distances between the cluster centers}\;
 \For{$( i = 1; i \leq K; i = i + 1 )$}
  {
   \For{$( j = i + 1; j \leq K; j = j + 1 )$}
    {
     $d[i][j] = d[j][i] = \| \bc_i - \bc_j \|^2$\;
    }
  }

 \CommentSty{Construct a $K \times K$ matrix $M$ in which row $i$ is a permutation of $1, 2, \ldots, K$ that
 represents the clusters in increasing order of distance of their centers from $\bc_i$}\;

 \For{$( i = 1; i \leq N'; i = i + 1 )$}
  {
   \CommentSty{Let $S_p$ be the cluster that $\bx_i$ was assigned to in the previous iteration}\;
   $p = m[i]$\;
   min\_dist $=$ prev\_dist $= \| \bx_i - \bc_p \|^2$\;

   \CommentSty{Update the nearest center if necessary}\;
   \For{$( j = 2; j \leq K; j = j + 1 )$}
    {
     $t = M[p][j]$\;
     \If{$d[p][t] \geq 4 \; prev\_dist$}
      {
       \CommentSty{There can be no other closer center. Stop checking}\;
       break\;
      }

     dist = $\| \bx_i - \bc_t \|^2$\;

     \If{$dist \leq min\_dist$}
      {
       \CommentSty{$\bc_t$ is closer to $\bx_i$ than $\bc_p$}\;
       min\_dist = dist\;
       $m[i] = t$;
      }
    }
  }
 \CommentSty{Recalculate the cluster centers}\;
 \For{$( k = 1; k \leq K; k = k + 1 )$}
  {
   \CommentSty{Calculate the new center $\bc_k$ as the weighted mean\\
    of points that are nearest to it}\;
   $\bc_k = {{\left( {\sum\limits_{m[i] = k} {w_i \bx_i } } \right)} \mathord{\left/
   {\vphantom {{\left( {\sum\limits_{m[i] = k} {w_i \bx_i } } \right)} {\sum\limits_{m[i] = k} {w_i } }}} \right.
   \kern-\nulldelimiterspace} {\sum\limits_{m[i] = k} {w_i } }}$\;
  }
}
\caption{Weighted Sort-Means Algorithm}
}
\end{algorithm}

\subsection{Initializing the K-Means Algorithm}
\label{sec_init}

It is well-known in the clustering literature that KM is quite sensitive to initialization. Adverse effects of improper initialization include:
\begin{inparaenum}[(i)]
 \item empty clusters (a.k.a.\ `dead units'),
 \item slower convergence, and
 \item a higher chance of getting stuck in bad local minima.
\end{inparaenum}
In this study, the following initialization schemes are investigated:

\begin{itemize}
 \item \textbf{Forgy (FGY)} \citep{Forgy65}: The cluster centers are chosen randomly from the data set. The complexity of this scheme is $\mathcal{O}(K)$.
 \item \textbf{Splitting (LBG)} \citep{Linde80}: The first center $\bc_1$ is chosen as the centroid of the data set. At iteration $i$ $(i \in \{ 1, 2, \ldots, \log_2K \} )$, each of the existing $2^{i-1}$ centers is split into two new centers by subtracting and adding a fixed perturbation vector $\boldsymbol \epsilon$, i.e.\ $\bc_j - \boldsymbol \epsilon$ and $\bc_j + \boldsymbol \epsilon$, $( j \in \{ 1, 2, \ldots, 2^{i-1} \} )$. These $2^i$ new centers are then refined using the KM algorithm. The complexity of this scheme is $\mathcal{O}(NK)$.
 \item \textbf{Minmax (MMX)} \citep{Hochbaum85,Gonzalez85,Katsavounidis94}: The first center $\bc_1$ is chosen randomly and the $i$-th $(i \in \{ 2, 3, \ldots, K \})$ center $\bc_i$ is chosen to be the point that has the largest minimum distance to the previously selected centers, i.e.\ $\bc_1, \bc_2, \ldots, \bc_{i-1}$. The complexity of this scheme is $\mathcal{O}(NK)$.
 \item \textbf{Density-based (DEN)} \citep{AlDaoud96}: The data space is partitioned uniformly into $M$ cells. From each of these cells, a number (that is proportional to the number of points in this cell) of centers is chosen randomly until $K$ centers are obtained. The complexity of this scheme is $\mathcal{O}(N)$. 
 \item \textbf{Maximum variance (VAR)} \citep{AlDaoud05}: The data set is sorted (in ascending or descending order) on the dimension that has the largest variance and then partitioned into $K$ groups along the same dimension. The centers are given by the data points that correspond to the medians of these $K$ groups. The complexity of this scheme is $\mathcal{O}(N \log{N})$.
 \item \textbf{Subset Farthest First (SFF)} \citep{Turnbull05}: One drawback of the MMX technique is that it tends to find the outliers in the data set. Using a smaller subset of size $2K \ln K$, the total number of outliers that MMX can find is reduced and thus the proportion of nonoutlier points obtained as centers is increased. The complexity of this scheme is $\mathcal{O}(K^2 \ln{K})$.
 \item \textbf{K-Means++ (KPP)} \citep{Arthur07}: The first center $\bc_1$ is chosen randomly and the $i$-th $(i \in \{ 2, 3, \ldots, K \})$ center $\bc_i$ is chosen to be $\bx' \in X$ with a probability of $\frac{{D\left( {\bx'} \right)^2 }}{{\sum\nolimits_{i = 1}^N {D(\bx_i )^2 } }}$, where $D(\bx)$ denotes the minimum distance from a point $\mathbf{x}$ to the previously selected centers.
\end{itemize}

In the remainder of this paper, these will be referred to as the \emph{generic initialization schemes} since they are applicable not only to color image data, but also to data with any dimensionality. Among these Forgy's scheme is the simplest and most commonly used one. However, as will be seen in the next section, this scheme often leads to poor clustering results. Note that there are numerous other initialization schemes described in the literature. These include methods based on hierarchical clustering \citep{Milligan80a}, genetic algorithms \citep{Babu93}, simulated annealing \citep{Babu94, Perim08}, multiscale data condensation \citep{Khan04}, and kd-trees \citep{Redmond07}. Other interesting methods include the global k-means method \citep{Likas03}, Kaufman and Rousseeuw's method \citep{Kaufman05}, and the ROBIN method \citep{AlHasan09}. Most of these schemes have quadratic or higher complexity in the number of points and therefore are not suitable for large data sets such as color image data.

\section{Experimental Results and Discussion}
\label{sec_exp}

\subsection{Image set and performance criteria}
\label{sec_criteria}

\begin{figure}[!ht]
\centering
 \subfigure[Airplane]{\includegraphics[width=0.2\columnwidth]{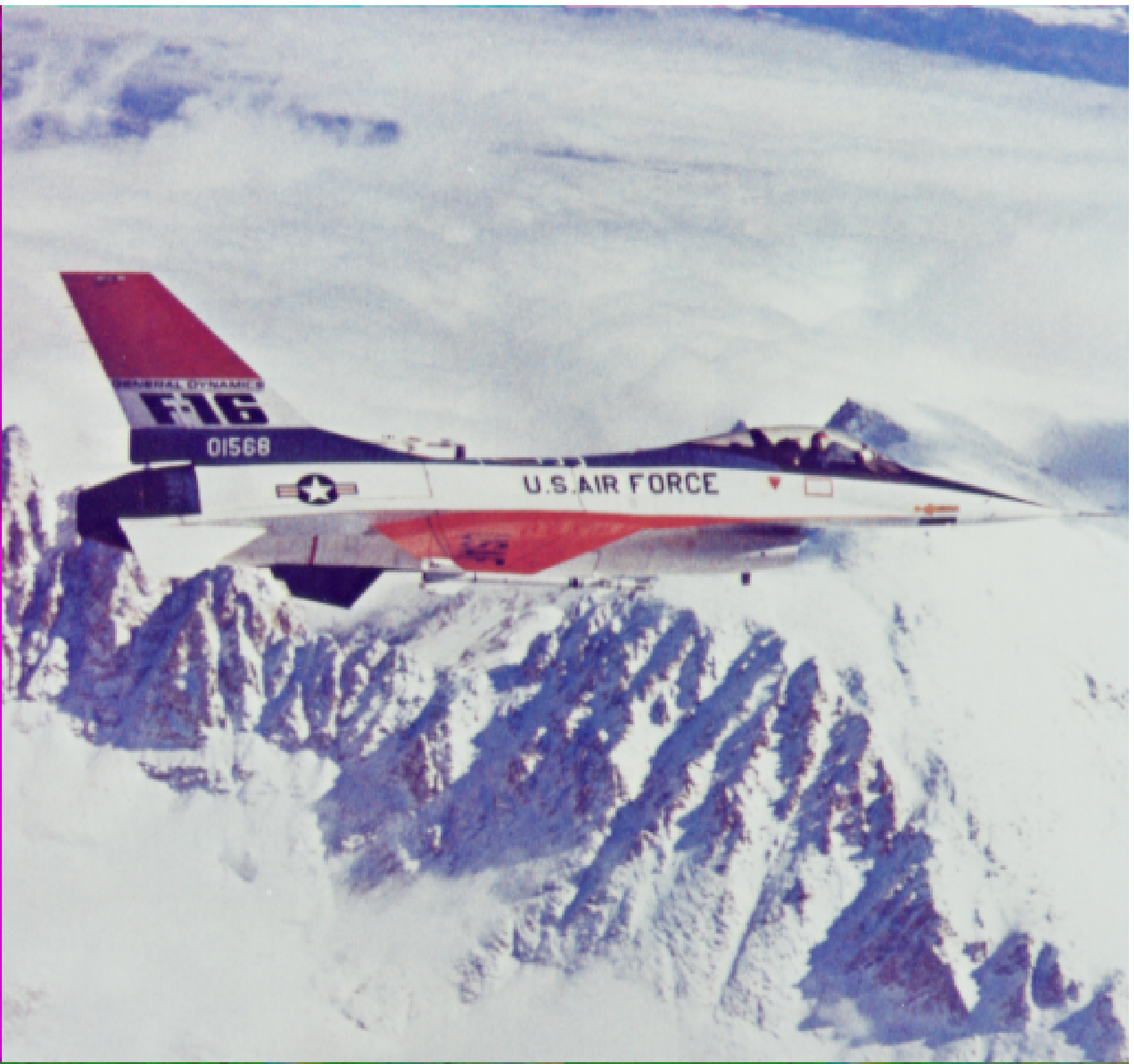}}
 \subfigure[Baboon]{\includegraphics[width=0.2\columnwidth]{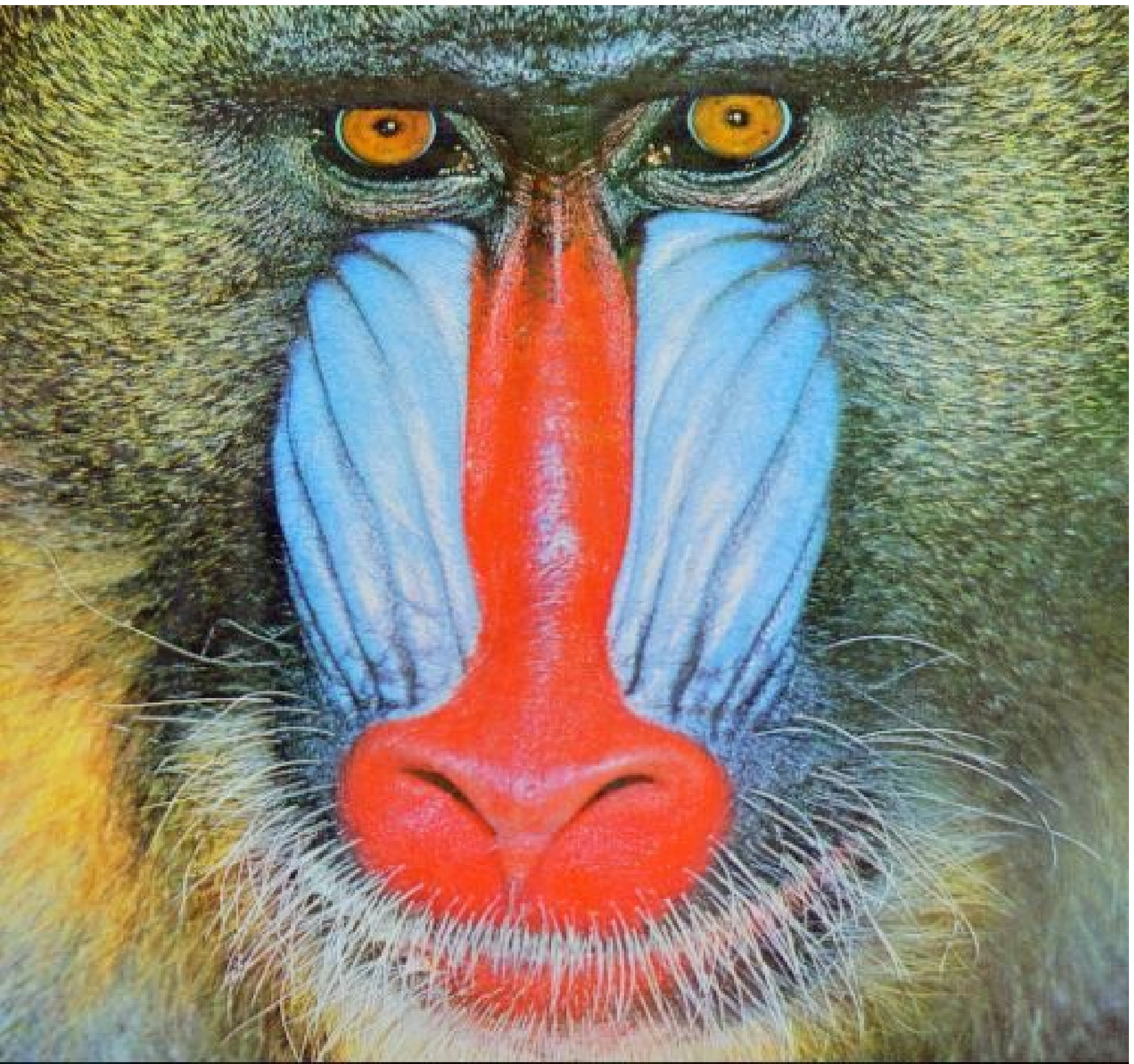}}
 \subfigure[Boats]{\includegraphics[width=0.2\columnwidth]{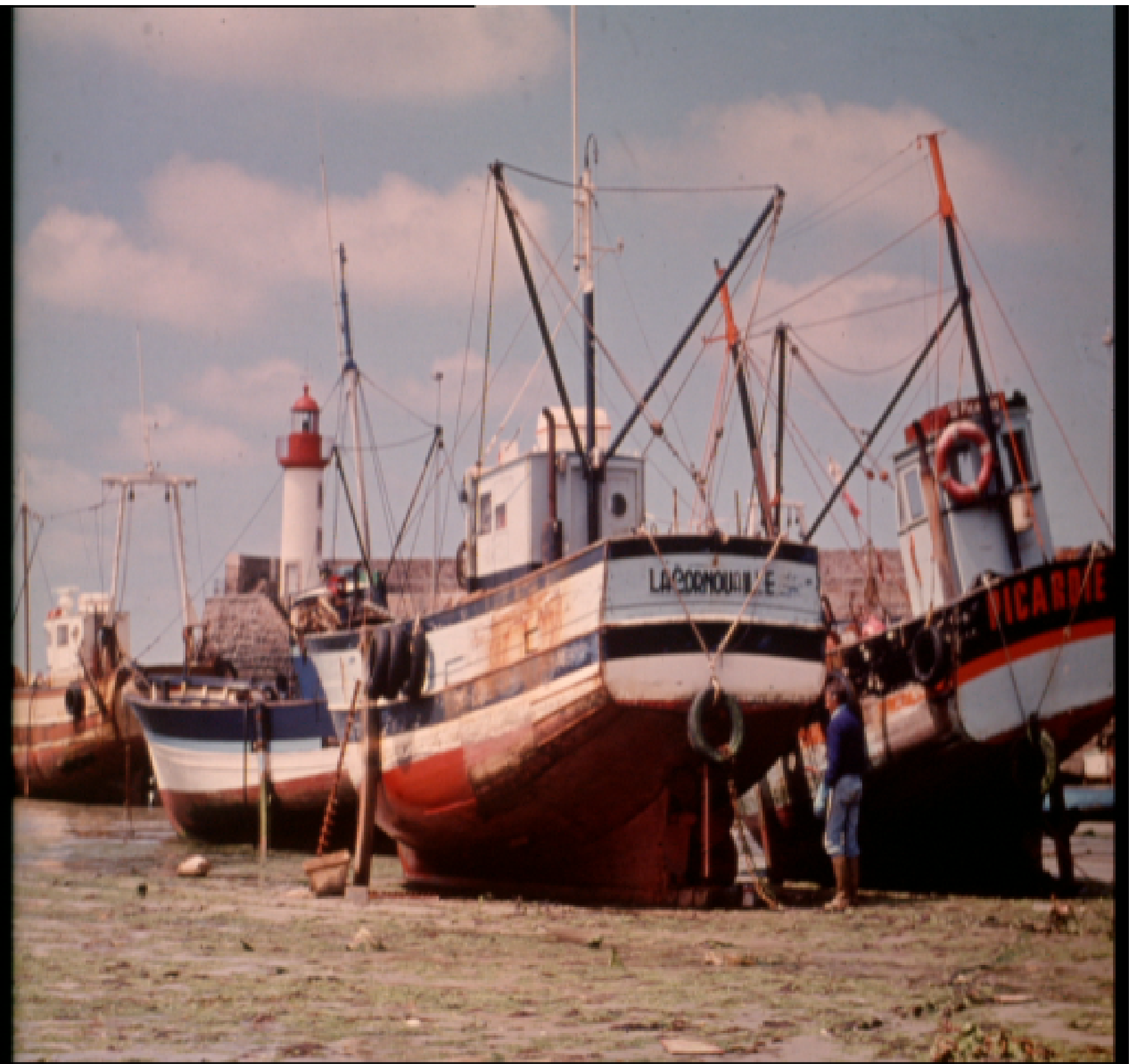}}
 \subfigure[Lenna]{\includegraphics[width=0.2\columnwidth]{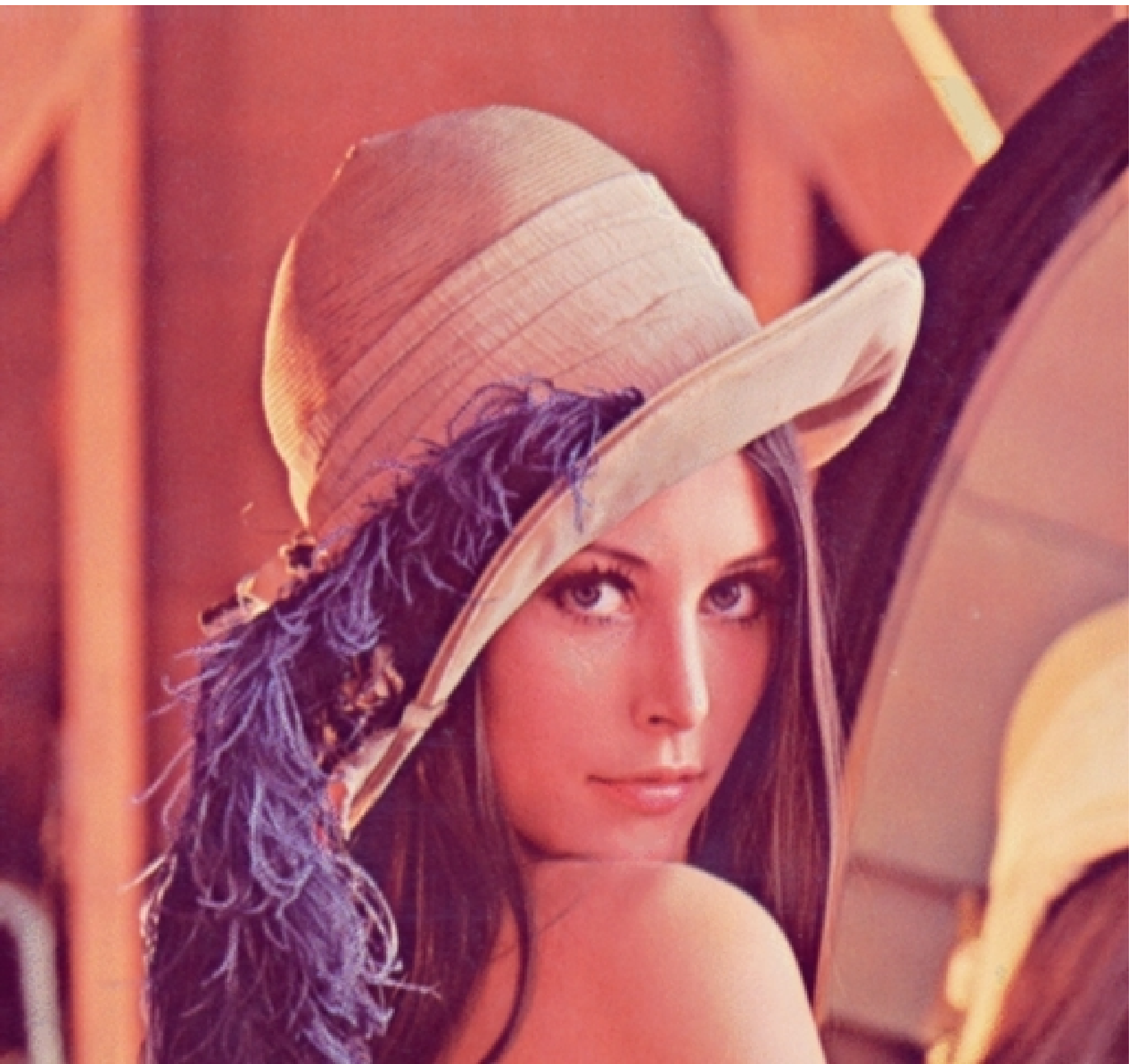}}
 \subfigure[Parrots]{\includegraphics[width=0.2\columnwidth]{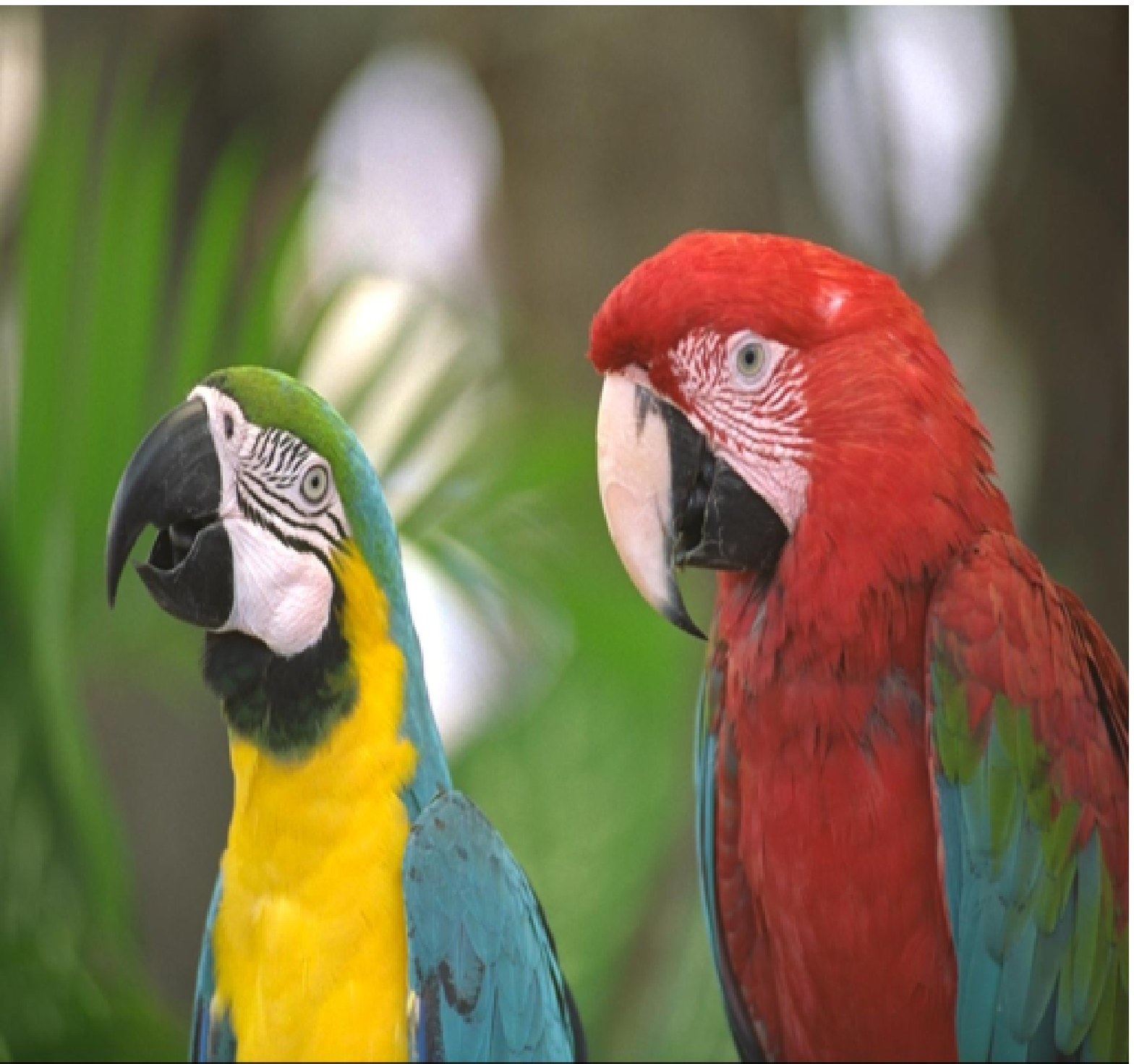}}
 \subfigure[Peppers]{\includegraphics[width=0.2\columnwidth]{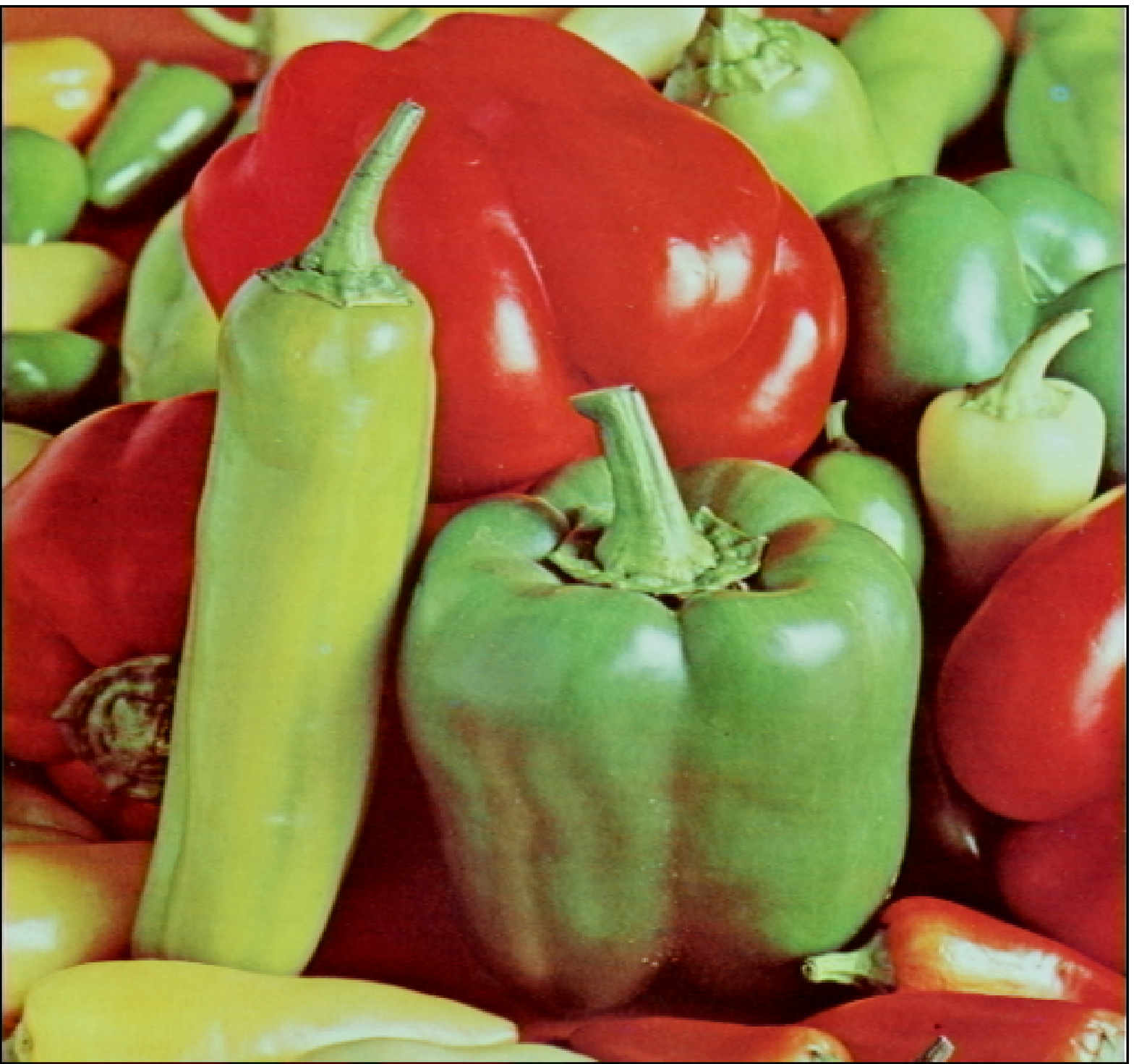}}
 \subfigure[Fish]{\includegraphics[width=0.2\columnwidth]{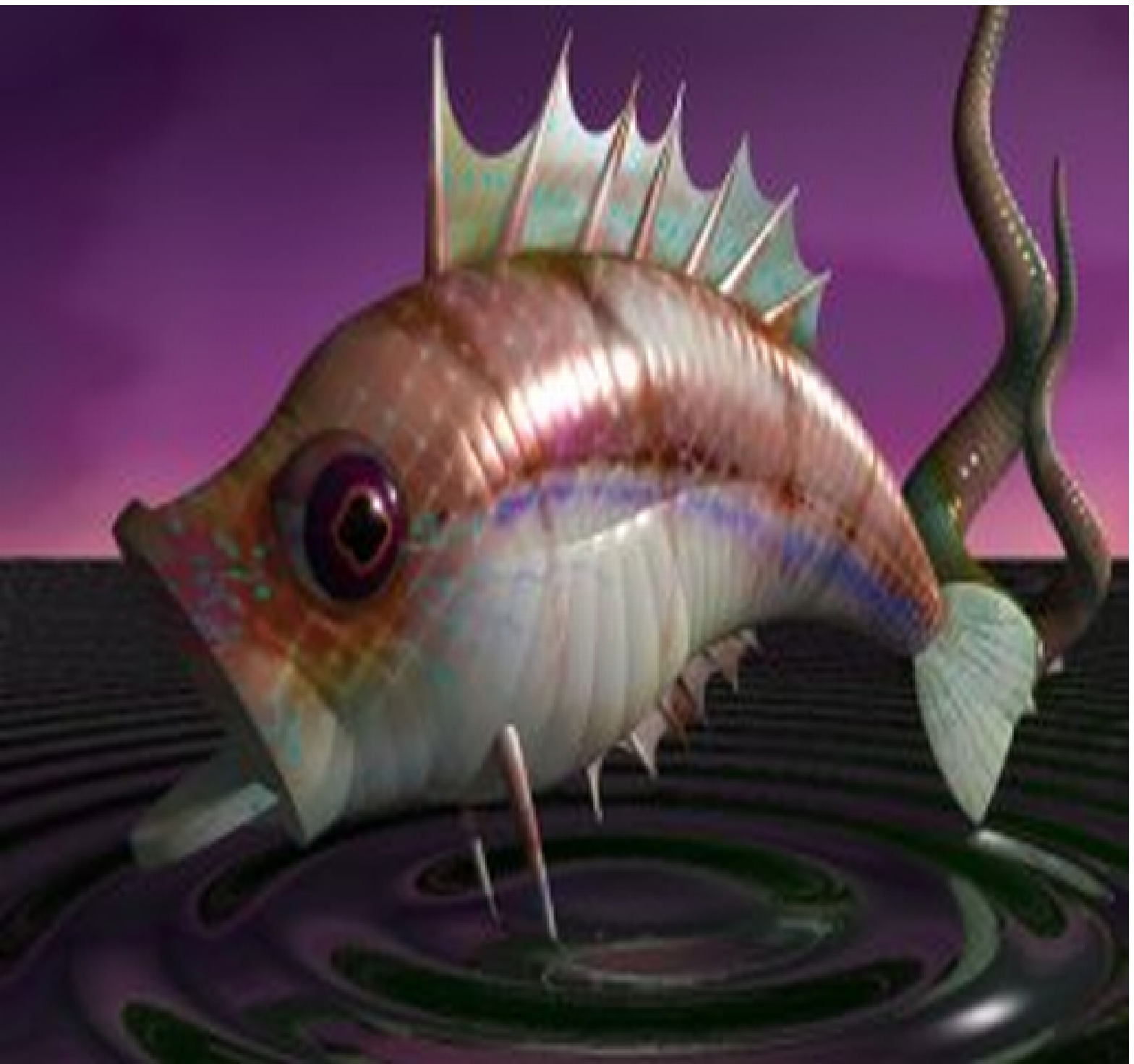}}
 \subfigure[Poolballs]{\includegraphics[width=0.2\columnwidth]{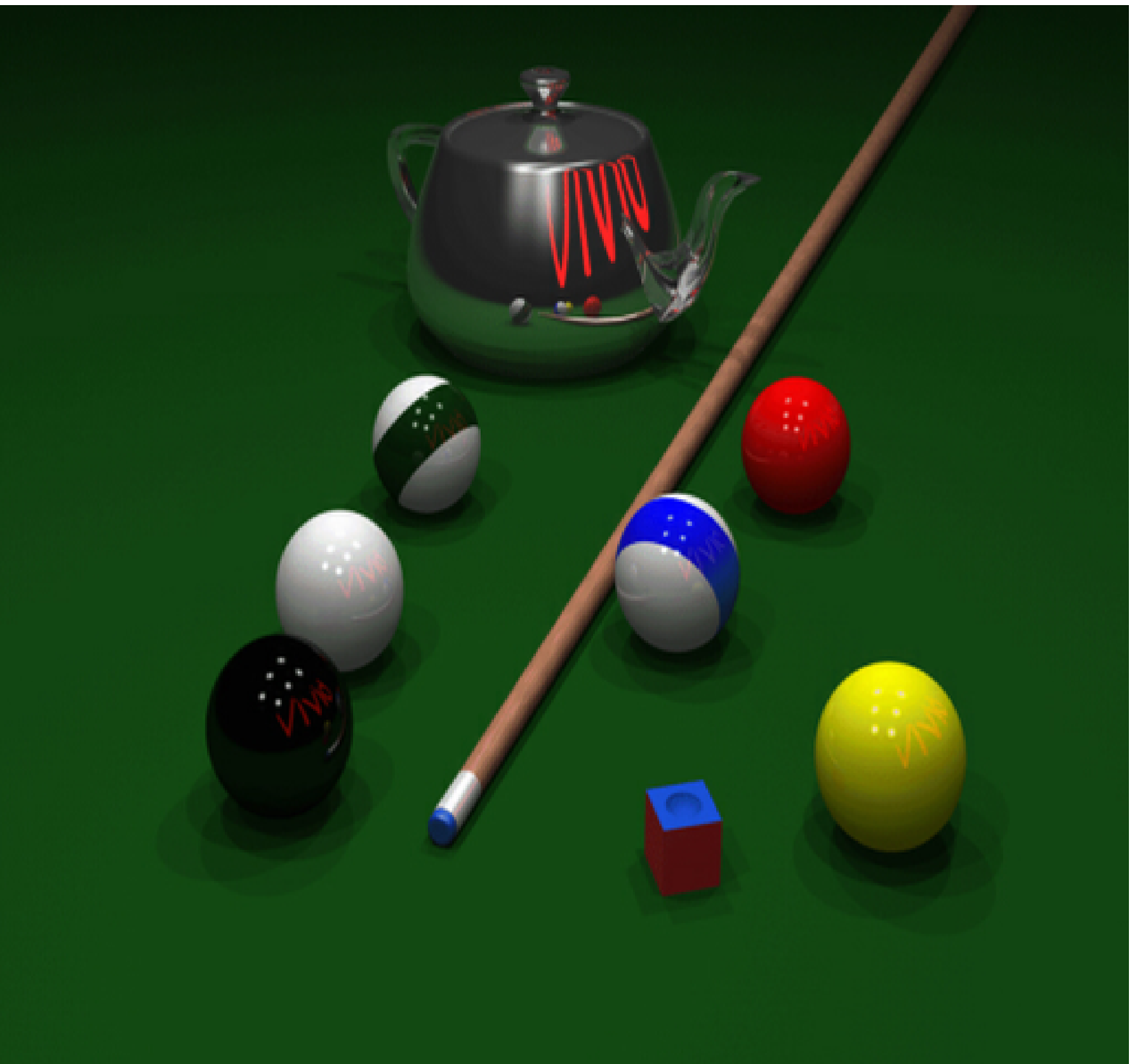}}
 \caption{Test images}
 \label{fig_test_images}
\end{figure}

The proposed method was tested on some of the most commonly used test images in the quantization literature (see Figure \ref{fig_test_images}). The natural images in the set include Airplane ($512 \times 512$, 77,041 (29\%) unique colors), Baboon ($512 \times 512$, 153,171 (58\%) unique colors), Boats ($787 \times 576$, 140,971 (31\%) unique colors), Lenna ($512 \times 512$, 148,279 (57\%) unique colors), Parrots ($1536 \times 1024$, 200,611 (13\%) unique colors), and Peppers ($512 \times 512$, 111,344 (42\%) unique colors). The synthetic images include Fish ($300 \times 200$, 28,170 (47\%) unique colors) and Poolballs ($510 \times 383$, 13,604 (7\%) unique colors).

The effectiveness of a quantization method was quantified by the Mean Squared Error (MSE) measure \citep{Brun02}:
\begin{equation}
	\mbox{MSE}\left( {{\bf X},{\bf \hat{X}}} \right) = \frac{1}{{HW}}\sum\limits_{h = 1}^H {\sum\limits_{w = 1}^W {\parallel {\bf x}(h,w) - {\bf \hat{x}}(h,w) \parallel}_2^2}
\end{equation}
where ${\bf X}$ and ${\bf \hat{X}}$ denote respectively the $H \times W$ original and quantized images in the RGB color space. MSE represents the average distortion with respect to the $L_2^2$ norm \eqref{eq_sse} and is the most commonly used evaluation measure in the quantization literature \citep{Brun02}. Note that the Peak Signal-to-Noise Ratio (PSNR) measure can be easily calculated from the MSE value:
\begin{equation}
 \mbox{PSNR} = 20 \log_{10} \left( {\frac{{255}}{{\sqrt {\mbox{MSE}} }}} \right).
\end{equation}
The efficiency of a quantization method was measured by CPU time in milliseconds, which includes the time required for both the palette generation and pixel mapping phases. In order to perform a fair comparison, the fast pixel mapping algorithm described in \citep{Hu08b} was used in quantization methods that lack an efficient pixel mapping phase. All of the programs were implemented in the C language, compiled with the gcc v4.2.4 compiler, and executed on an Intel Xeon E5520 2.26GHz machine. The time figures were averaged over 100 runs.

\subsection{Efficiency comparison between WSM and KM}
\label{sec_wsm_km}

In this subsection, the computational efficiency of WSM is compared to that of KM. In order to ensure fairness in the comparisons, both algorithms were initialized with the same randomly chosen centers and terminated after 20 iterations. Table \ref{tab_ndc_time} gives the Number of Distance Calculations (NDC) per pixel and computational times for $K=\{32, 64, 128, 256\}$ on the test images. Note that for KM, NDC always equals the palette size $K$ since the nearest neighbor search involves a full search of the palette for each input pixel. In contrast, WSM requires, on the average, 8--16 times fewer distance calculations, which is due to the intelligent use of the triangle inequality that avoids many calculations once the cluster centers stabilize after a few iterations. Most KM acceleration methods incur overhead to create and update auxiliary data structures. This means that speed up when compared to KM is less in CPU time than in NDC \citep{Elkan03}. Table \ref{tab_ndc_time} shows that this is not the case for WSM since it exploits the color redundancy in the original images by reducing the amount of data before the clustering phase. It can be seen that WSM is actually about 12--20 times faster than KM. Note that the speed up for a particular image is inversely proportional to the number of unique colors in the image. Therefore, the most significant computational savings are observed on images with relatively few unique colors such as Parrots (13\% unique colors) and Poolballs (7\% unique colors).
\par
\begin{figure}[!ht]
\centering
\includegraphics[width=0.72\columnwidth,draft=false]{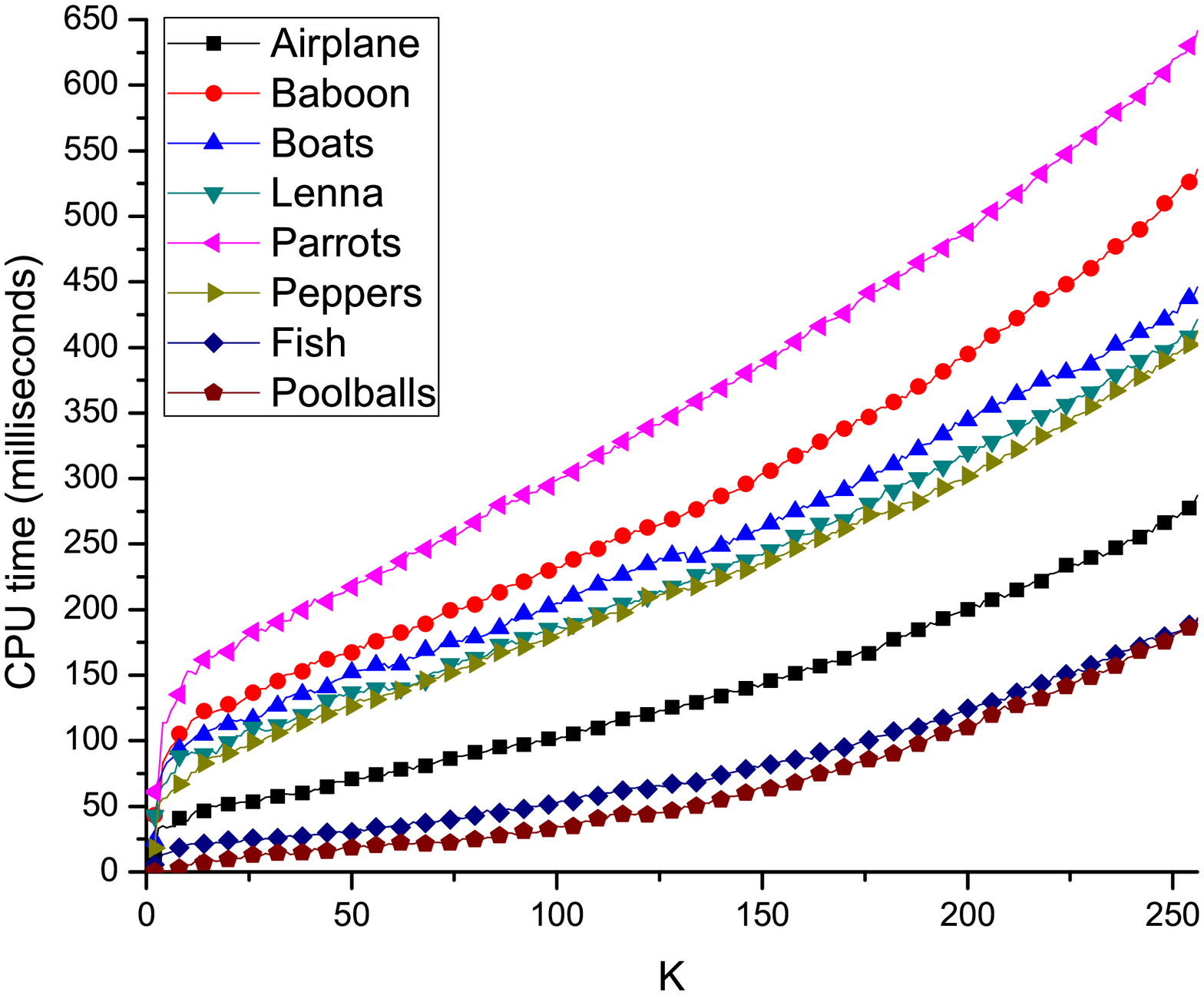}
\caption { \label{fig_graph} CPU time for WSM for $K=\left\{ 2, 3, \ldots, 256 \right\}$ }
\end{figure}

Figure \ref{fig_graph} illustrates the scaling behavior of WSM with respect to $K$. It can be seen that, in contrast to KM, the complexity of WSM is sublinear in $K$. For example, on the Parrots image, increasing $K$ from 16 to 256, results in only about 3.9 fold increase in the computational time (164 ms vs.\ 642 ms).

\subsection{Comparison of WSM against other quantization methods}
\label{sec_comp}

The WSM algorithm was compared to some of the well-known quantization methods in the literature:

\begin{itemize}
 \item \textbf{Median-cut (MC)} \citep{Heckbert82}: This method starts by building a $32 \times 32 \times 32$ color histogram that contains the original pixel values reduced to 5 bits per channel by uniform quantization. This histogram volume is then recursively split into smaller boxes until $K$ boxes are obtained. At each step, the box that contains the largest number of pixels is split along the longest axis at the median point, so that the resulting subboxes each contain approximately the same number of pixels. The centroids of the final $K$ boxes are taken as the color palette.

 \item \textbf{Otto's method (OTT)} \citep{Otto87}: This method is similar to MC with two exceptions: no uniform quantization is performed and at each step the box that gives the maximum reduction in the total squared deviation is split. The split axis and split point are determined by exhaustive search.

 \item \textbf{Octree (OCT)} \citep{Gervautz88}: This two-phase method first builds an octree (a tree data structure in which each internal node has up to eight children) that represents the color distribution of the input image and then, starting from the bottom of the tree, prunes the tree by merging its nodes until $K$ colors are obtained. In the experiments, the tree depth was limited to 6.

 \item \textbf{Variance-based method (WAN)} \citep{Wan90}: This method is similar to MC with the exception that at each step the box with the largest weighted variance (squared error) is split along the major (principal) axis at the point that minimizes the marginal squared error.

 \item \textbf{Greedy orthogonal bipartitioning (WU)} \citep{Wu91}: This method is similar to WAN with the exception that at each step the box with the largest weighted variance is split along the axis that minimizes the sum of the variances on both sides.

 \item \textbf{Binary splitting method (BS)} \citep{Orchard91}: This method is similar to WAN with two exceptions: no uniform quantization is performed and at each step the box with the largest eigenvalue is split along the major axis at the mean point.

 \item \textbf{Neu-quant (NEU)} \citep{Dekker94}: This method utilizes a one-dimensional self-organizing map (Kohonen neural network) with 256 neurons. A random subset of $N/f$ pixels is used in the training phase and the final weights of the neurons are taken as the color palette. In the experiments, the highest quality configuration, i.e.\ $f = 1$, was used.

 \item \textbf{Modified minmax (MMM)} \citep{Xiang97}: This method chooses the first center $\bc_1$ arbitrarily from the data set and the $i$-th center $\bc_i$ ($i = 2, 3, \ldots, K$) is chosen to be the point that has the largest minimum weighted $L_2^2$ distance (the weights for the red, green, and blue channels are taken as 0.5, 1.0, and 0.25, respectively) to the previously selected centers, i.e.\ $\bc_1, \bc_2, \ldots, \bc_{i-1}$. Each of these initial centers is then recalculated as the mean of the points assigned to it.

 \item \textbf{Split \& Merge (SAM)} \citep{Brun00}: This two-phase method first partitions the color space uniformly into $B$ partitions. This initial set of $B$ clusters is represented as an adjacency graph. In the second phase, $(B - K)$ merge operations are performed to obtain the final $K$ clusters. At each step of the second phase, the pair of clusters with the minimum joint quantization error are merged. In the experiments, the initial number of clusters was set to $B = 20 K$. 

 \item \textbf{Fuzzy c-Means with partition index maximization (PIM)} \citep{Ozdemir02}: Fuzzy c-Means (FCM) \citep{Bezdek81} is a generalization of KM in which points can belong to more than one cluster. The algorithm involves the minimization of the functional $J_q(U,V) = \sum\nolimits_{i = 1}^N {\sum\nolimits_{k = 1}^K {u_{ik}^q \left\| {\bx_i - \bv_k } \right\|_2^2 } }$ with respect to $U$ (a fuzzy $K$-partition of the data set) and $V$ (a set of prototypes -- cluster centers). The parameter $q$ controls the fuzziness of the resulting clusters. At each iteration, the membership matrix $U$ is updated by
 $u_{ik}  = \left( {\sum\nolimits_{j = 1}^K {\left( {{{\left\| {\bx_i - \bv_k } \right\|_2 } \mathord{\left/
 {\vphantom {{\left\| {\bx_i - \bv_k } \right\|_2 } {\left\| {\bx_i - \bv_j } \right\|_2 }}} \right.
 \kern-\nulldelimiterspace} {\left\| {\bx_i - \bv_j } \right\|_2 }}} \right)^{{2 \mathord{\left/
 {\vphantom {2 {(q - 1)}}} \right. \kern-\nulldelimiterspace} {(q - 1)}}} } } \right)^{ - 1}$,
which is followed by the update of the prototype matrix $V$ by
$\bv_k  = {{\left( {\sum\nolimits_{i = 1}^N {u^q _{ik} \bx_i } } \right)} \mathord{\left/ {\vphantom {{\left( {\sum\nolimits_{i = 1}^N {u^q _{ik} \bx_i } } \right)} {\left( {\sum\nolimits_{i = 1}^N {u^q _{ik} } } \right)}}} \right. \kern-\nulldelimiterspace} {\left( {\sum\nolimits_{i = 1}^N {u^q _{ik} } } \right)}}$.
A n\"{a}ive implementation of the FCM algorithm has a complexity of $\mathcal{O}(NK^2)$ per iteration, which is quadratic in the number of clusters. In the experiments, a linear complexity formulation, i.e.\ $\mathcal{O}(NK)$, described in \citep{Kolen02} was used and the fuzziness parameter was set to $q=2$ as commonly seen in the fuzzy clustering literature \citep{Gan07}. PIM is an extension of FCM in which the functional to be minimized incorporates a cluster validity measure called the 'partition index' (PI). This index measures how well a point $\bx_i$ has been classified and is defined as $P_i  = \sum\nolimits_{k = 1}^K {u^q _{ik} }$. The FCM functional can be modified to incorporate PI as follows: $J^\alpha_q(U,V) = \sum\nolimits_{i = 1}^N {\sum\nolimits_{k = 1}^K {u_{ik}^q \left\| {\bx_i - \bv_k } \right\|_2^2 } } - \alpha \sum\nolimits_{i = 1}^N {P_i}$. The parameter $\alpha$ controls the weight of the second term. The procedure that minimizes $J^\alpha_q(U,V)$ is identical to the one used in FCM except for the membership matrix update equation:
 $u_{ik}  = \left( {\sum\nolimits_{j = 1}^K {\left[ {{{\left( {\left\| {\bx_i  - \bv_k } \right\|_2  - \alpha } \right)} \mathord{\left/
 {\vphantom {{\left( {\left\| {\bx_i  - \bv_k } \right\|_2  - \alpha } \right)} {\left( {\left\| {\bx_i - \bv_j } \right\|_2  - \alpha } \right)}}} \right. \kern-\nulldelimiterspace} {\left( {\left\| {\bx_i - \bv_j } \right\|_2  - \alpha} \right)}}} \right]^{{2 \mathord{\left/
 {\vphantom {2 {(q - 1)}}} \right. \kern-\nulldelimiterspace} {(q - 1)}}} } } \right)^{- 1}$.
An adaptive method to determine the value of $\alpha$ is to set it to a fraction $0 \leq \delta < 0.5$ of the distance between the nearest two centers, i.e.\ $\alpha = \delta \min\limits_{i \ne j} {\left\| {\bv_i - \bv_j } \right\|_2^2}$. Following \citep{Ozdemir02}, the fraction value was set to $\delta = 0.4$.

 \item \textbf{Competitive learning clustering (ADU)} \citep{Celebi09}: This method is an adaptation of Uchiyama and Arbib's Adaptive Distributing Units (ADU) algorithm \citep{Uchiyama94a} to color quantization. ADU is a competitive learning algorithm in which units compete to represent the input point presented in each iteration. The winner is then rewarded by moving it closer to the input point at a rate of $\gamma$ (the learning rate). The procedure starts with a single unit whose center is given by the centroid of the input points. New units are added by splitting existing units that reach the threshold number of wins $\theta$ until the number of units reaches $K$. Following \citep{Celebi09}, the algorithm parameters were set to $\theta = 400\sqrt{K}$, $t_{max} = (2K - 3)\theta$, and $\gamma = 0.015$.
\end{itemize}

Fourteen variants of WSM each with a different initialization scheme were implemented. These include variants that utilize the generic initialization schemes discussed in \S \ref{sec_init}, namely WSM-FGY, WSM-LBG, WSM-MMX, WSM-DEN, WSM-VAR, WSM-SFF, and WSM-KPP, as well as variants that use the abovementioned preclustering methods as initializers, i.e.\ WSM-MC, WSM-OTT, WSM-OCT, WSM-WAN, WSM-WU, WSM-BS, and WSM-SAM. Each variant was executed until it converged. Convergence was determined by the following commonly used criterion \citep{Linde80}: $ {{\left( {\mbox{SSE}_{i - 1} - \mbox{SSE}_i} \right)} \mathord{\left/ {\vphantom {{\left( {\mbox{SSE}_{i - 1} - \mbox{SSE}_i} \right)} {\mbox{SSE}_i}}} \right. \kern-\nulldelimiterspace} {\mbox{SSE}_i}} \leq \varepsilon $, where $\mbox{SSE}_i$ denotes the SSE \eqref{eq_sse} value at the end of the $i$-th iteration. The convergence threshold was set to $\varepsilon = 0.001$.
\par
Tables \ref{tab_mse_32}-\ref{tab_mse_256} compare the effectiveness of the methods for 32, 64, 128, and 256 colors, respectively on the test images. The best (lowest) error values are shown in \textbf{bold}. Similarly, Tables \ref{tab_time_32}-\ref{tab_time_256} give the efficiency comparison of the methods. In addition, for each $K$ value, the methods are first ranked based on their MSE values for each image. These ranks are then averaged over all test images. The same is done for the CPU time values. Table \ref{tab_mse_time_rank} gives the mean MSE and CPU time ranks of the methods. The last column gives the overall mean ranks with the assumption that both criteria have equal importance. Note that the best possible rank is 1. The following observations are in order:

\begin{itemize}
\renewcommand{\labelitemi}{$\triangleright$}
 \item In general, the postclustering methods are more effective but less efficient than the preclustering methods.
 \item The most effective preclustering methods are BS, OTT, and WU. The least effective ones are MC, WAN, and MMM.
 \item The most effective postclustering methods are WSM-WU, WSM-BS, WSM-WAN, WSM-SAM, and WSM-LBG. Note that two of these methods, namely WSM-WAN and WSM-SAM, utilize initialization methods that are quite ineffective by themselves. (The MSE ranks of WAN and SAM are 24.00 and 20.19, respectively.) The least effective postclustering methods are PIM, NEU, and WSM-MMX.
 \item Preclustering methods are generally more effective and efficient (especially when $K$ is small) initializers when compared to the generic schemes. This was expected since the former methods are designed to exploit the peculiarities of color image data such as limited range and sparsity. Therefore, they are particularly suited for time-constrained applications such as color based retrieval from large image databases, where images are often reduced to a few colors prior to the similarity calculations \citep{Deng01b}.
 \item In general, WSM-WU is the best method. This method is not only the overall most effective method, but also the most efficient postclustering method. In each case, it obtains one of the lowest MSE values within a fraction of a second.
 \item In general, the fastest method is MC, which is followed by WU, WAN, and SAM. The slowest methods are PIM, WSM-LBG, MMM, ADU, NEU, and BS.
\end{itemize}

Table \ref{tab_iters} gives the number of iterations that each WSM variant requires until reaching convergence. As before, for each $K$ value, the methods are first ranked based on their iteration counts for each image. These ranks are then averaged over all test images. Table \ref{tab_iters_rank} gives the results, with the last column representing the overall mean ranks. The correlation coefficient between this column and the overall mean MSE ranks (column 6 of Table \ref{tab_mse_time_rank}) is 0.882, which indicates that WSM often takes longer to converge when initialized by an ineffective preclustering method. Interestingly, despite the fact that it converges the fastest, WSM-LBG is one of the slowest quantization methods because of its costly initialization phase, i.e.\ the LBG algorithm. In fact, the correlation coefficient between the overall mean iteration count ranks and the overall mean CPU time ranks (column 11 of Table \ref{tab_mse_time_rank}) is 0.034, which indicates a lack of correlation between the number of iterations and computational speed.
\par
It should be noted that initializing a postclustering method such as KM (or WSM) using the output (color palette) generated by a preclustering method is not a new idea. Numerous early studies \citep{Heckbert82,Braudaway87,Equitz89,Wan90,Gentile90,Orchard91,Goldberg91,Pei95} investigated this particular two-phase quantization scheme and concluded that slight improvements (reductions) in the MSE due to the use of KM is largely offset by the dramatic increase in the computational time. Table \ref{tab_mse_improvement} gives the percent MSE improvements obtained by refining the outputs generated by the preclustering methods  using WSM. For example, when $K=32$, WSM-MC obtains, on the average, 42\% lower MSE when compared to MC on the test images. It can be seen that WSM improves the MSE values by an average of 18--50\%. When combined with its significant computational efficiency (see \S \ref{sec_wsm_km}), these improvements show that the conclusions made for KM in the abovementioned studies are not valid for WSM. The correlation coefficient between the mean percent MSE improvement values (last column of Table \ref{tab_mse_improvement}) and the overall mean MSE ranks (column 6 of Table \ref{tab_mse_time_rank}) is $0.988$, which indicates that WSM is much more likely to obtain a significant MSE improvement when initialized by an ineffective preclustering method such as MC or WAN. This is not surprising given that such ineffective methods generate outputs that are likely to be far from a local minimum and hence WSM can significantly improve upon their results. Nevertheless, it can be said that WSM benefits even highly effective preclustering methods such as BS, OTT, and WU.
\par
\begin{figure}[!ht]
\centering
 \subfigure[Original]{\includegraphics[width=0.2\columnwidth]{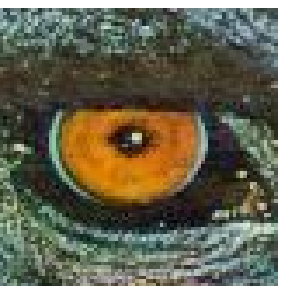}}
 \\
 \subfigure[PIM out]{\includegraphics[width=0.2\columnwidth]{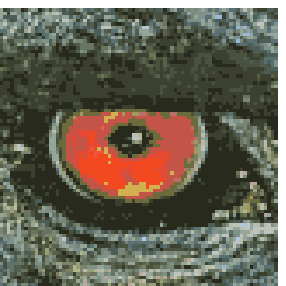}}
 \subfigure[PIM err]{\includegraphics[width=0.2\columnwidth]{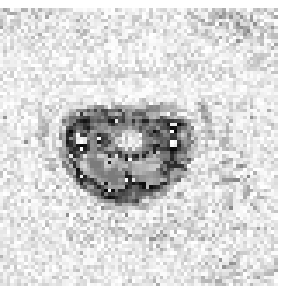}}
 \subfigure[NEU out]{\includegraphics[width=0.2\columnwidth]{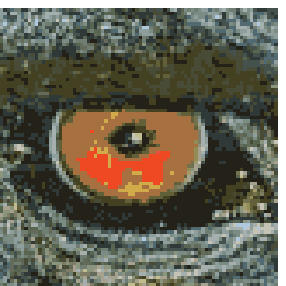}}
 \subfigure[NEU err]{\includegraphics[width=0.2\columnwidth]{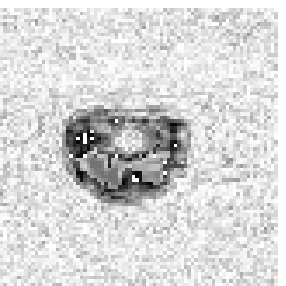}}
 \\
 \subfigure[BS out]{\includegraphics[width=0.2\columnwidth]{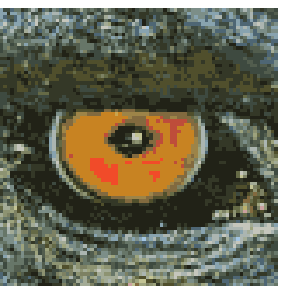}}
 \subfigure[BS err]{\includegraphics[width=0.2\columnwidth]{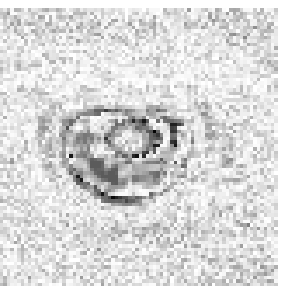}}
 \subfigure[WU out]{\includegraphics[width=0.2\columnwidth]{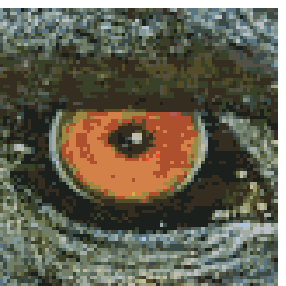}}
 \subfigure[WU err]{\includegraphics[width=0.2\columnwidth]{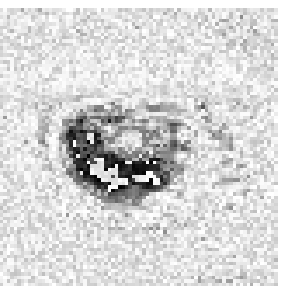}}
 \\
 \subfigure[WSM-BS out]{\includegraphics[width=0.2\columnwidth]{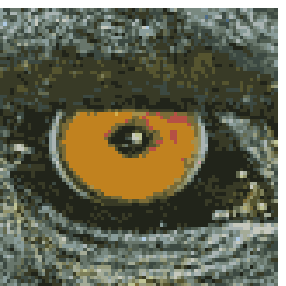}}
 \subfigure[WSM-BS err]{\includegraphics[width=0.2\columnwidth]{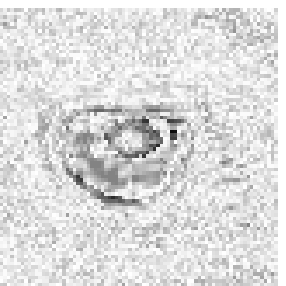}}
 \subfigure[WSM-WU out]{\includegraphics[width=0.2\columnwidth]{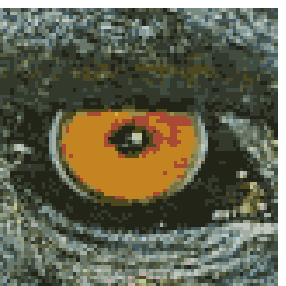}}
 \subfigure[WSM-WU err]{\includegraphics[width=0.2\columnwidth]{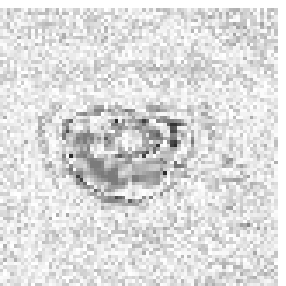}}
 \caption{Sample quantization results for the Baboon image ($K=32$)}
 \label{fig_baboon}
\end{figure}

Figure \ref{fig_baboon} shows sample quantization results and the corresponding error images for a close-up part of the Baboon image. The error image for a particular quantization method was obtained by taking the pixelwise absolute difference between the original and quantized images. In order to obtain a better visualization, pixel values of the error images were multiplied by 4 and then negated. It can be seen that PIM, NEU, BS, and WU are unable to represent the color distribution of the sclera (yellow part of the eye). This is because, this region is a relatively small part of the face and therefore, despite its visual significance, it is assigned representative colors that are derived from larger regions with different color distributions, e.g.\ the red nose. In contrast, WSM variants, i.e.\ WSM-BS and WSM-WU, perform significantly better in allocating representative colors to the sclera, resulting in cleaner error images.
\par
\begin{figure}[!ht]
\centering
 \subfigure[Original]{\includegraphics[width=0.2\columnwidth]{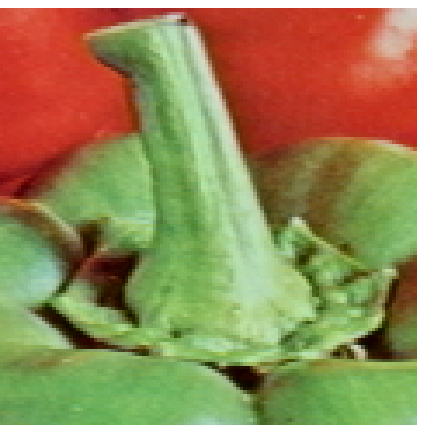}}
 \\
 \subfigure[PIM out]{\includegraphics[width=0.2\columnwidth]{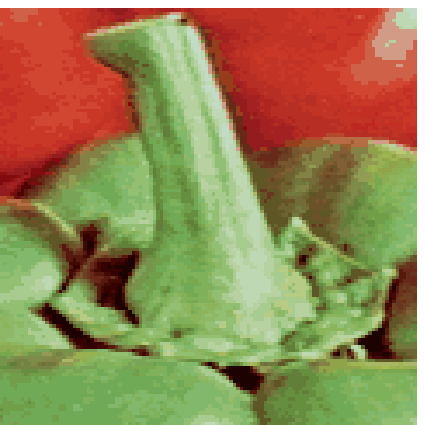}}
 \subfigure[PIM err]{\includegraphics[width=0.2\columnwidth]{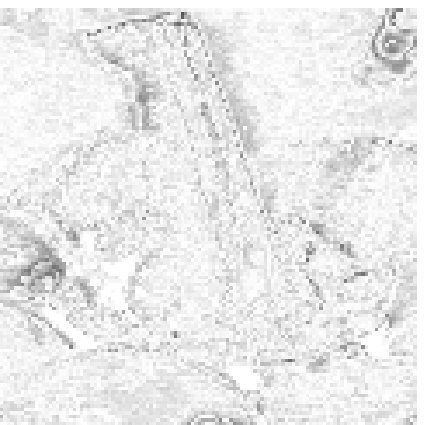}}
 \subfigure[NEU out]{\includegraphics[width=0.2\columnwidth]{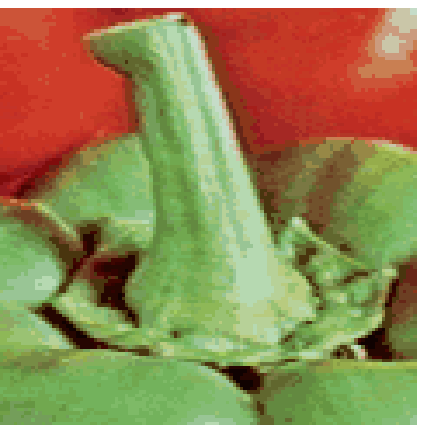}}
 \subfigure[NEU err]{\includegraphics[width=0.2\columnwidth]{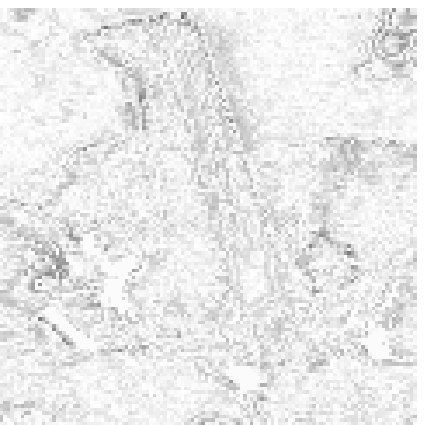}}
 \\
 \subfigure[MC out]{\includegraphics[width=0.2\columnwidth]{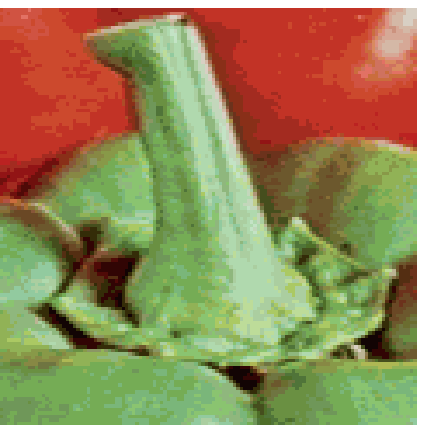}}
 \subfigure[MC err]{\includegraphics[width=0.2\columnwidth]{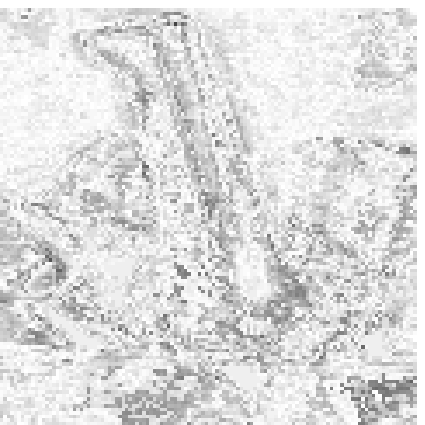}}
 \subfigure[WAN out]{\includegraphics[width=0.2\columnwidth]{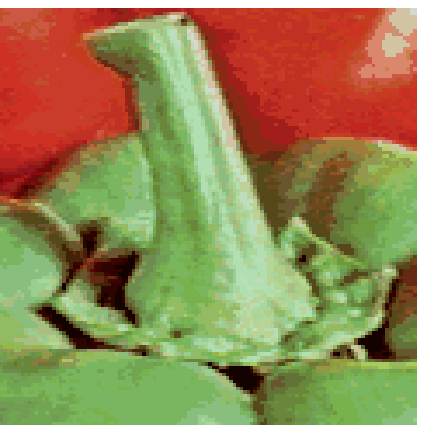}}
 \subfigure[WAN err]{\includegraphics[width=0.2\columnwidth]{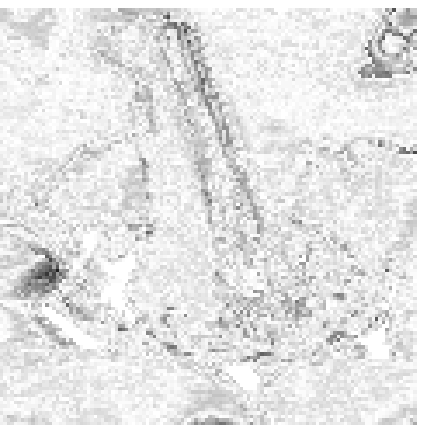}}
 \\
 \subfigure[WSM-MC out]{\includegraphics[width=0.2\columnwidth]{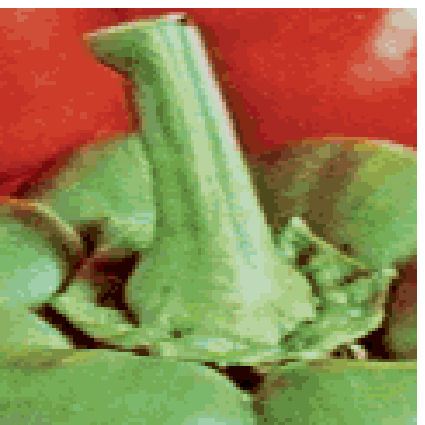}}
 \subfigure[WSM-MC err]{\includegraphics[width=0.2\columnwidth]{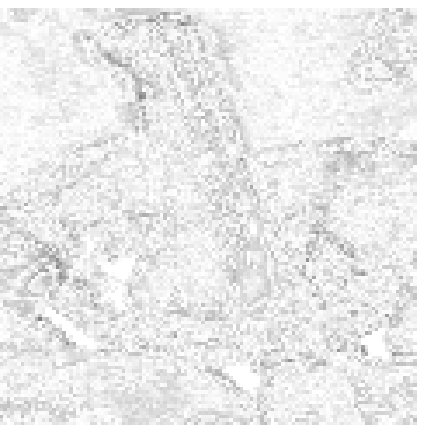}}
 \subfigure[WSM-WAN out]{\includegraphics[width=0.2\columnwidth]{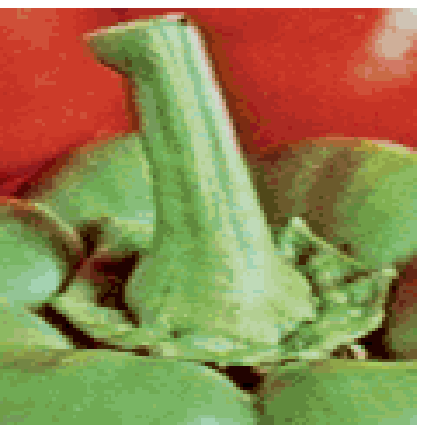}}
 \subfigure[WSM-WAN err]{\includegraphics[width=0.2\columnwidth]{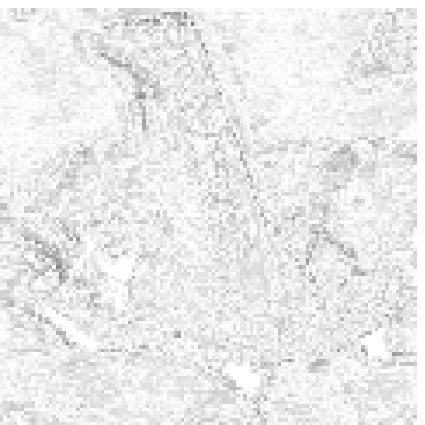}}
 \caption{Sample quantization results for the Peppers image ($K=64$)}
 \label{fig_peppers}
\end{figure}

Figure \ref{fig_peppers} shows sample quantization results and the corresponding error images for a close-up part of the Peppers image. It can be seen that PIM, NEU, MC, and WAN are particularly ineffective around the edges. On the other hand, WSM variants, i.e.\ WSM-MC and WSM-WAN, are significantly better in representing this edgy region. Once again, the refinement due to WSM is remarkable for the preclustering methods WAN and, in particular, MC.
\par
We should also mention a recent study by Chen \emph{et al.} that involves color quantization and the KM algorithm \citep{Chen08}. In their method, the input image is first quantized uniformly in the Hue-Saturation-Value (HSV) color space \citep{Smith78} to obtain a color histogram with $30 \times 7 \times 7$ bins and a grayscale one with $8$ bins. Initial cluster centers are then determined from each histogram using a modified MMX procedure that selects a maximum of $10$ centers using an empirically determined distance threshold. Finally, the histograms are jointly clustered using the KM algorithm and the resulting image is post-processed to eliminate small regions. To summarize, this method aims to partition the input image into a number of homogeneously colored regions using an image-independent quantization scheme and histogram clustering. In contrast, the proposed methods aim to reduce the number of colors in the input image to a predefined number using an image-dependent scheme. However, both approaches involve KM clustering on histogram data.

\begin{table}[ht]
\linespread{1}
\centering
\scriptsize
{
\caption{ \label{tab_ndc_time} NDC and CPU time comparison between WSM and KM}
\begin{tabular}{|c|c|c|c|c|c|c|c|c|c|c||c|}
\hline
K & Criterion & Method & AIR & BBN & BTS & LEN & PAR & PEP & FSH & PLB & Mean\\
\hline
\hline
\multirow{6}{*}{32} & \multirow{3}{*}{NDC} & KM & 32 & 32 & 32 & 32 & 32 & 32 & 32 & 32 & 32\\
& & WSM & 3.20 & 4.02 & 4.00 & 3.00 & 4.09 & 4.04 & 4.37 & 5.14 & 3.98\\
& & KM:WSM & 10.00 & 7.95 & 8.00 & 10.67 & 7.83 & 7.91 & 7.33 & 6.23 & \textbf{8.24}\\
\cline{2-12}
& \multirow{3}{*}{Time} & KM & 802 & 899 & 1415 & 858 & 4718 & 828 & 192 & 551 & 1283\\
& & WSM & 58 & 153 & 134 & 118 & 207 & 109 & 31 & 20 & 104\\
& & KM:WSM & 13.64 & 5.87 & 10.52 & 7.24 & 22.76 & 7.59 & 6.08 & 26.90 & \textbf{12.58}\\
\hline
\multirow{6}{*}{64} & \multirow{3}{*}{NDC} & KM & 64 & 64 & 64 & 64 & 64 & 64 & 64 & 64 & 64\\
& & WSM & 4.62 & 5.73 & 5.69 & 4.45 & 5.89 & 6.18 & 6.25 & 6.66 & 5.68\\
& & KM:WSM & 13.86 & 11.17 & 11.24 & 14.39 & 10.86 & 10.36 & 10.24 & 9.61 & \textbf{11.47}\\
\cline{2-12}
& \multirow{3}{*}{Time} & KM & 1540 & 1671 & 2708 & 1630 & 9392 & 1600 & 367 & 1069 & 2497\\
& & WSM & 81 & 191 & 172 & 174 & 259 & 150 & 41 & 27 & 137\\
& & KM:WSM & 18.92 & 8.74 & 15.66 & 9.34 & 36.20 & 10.66 & 8.84 & 38.76 & \textbf{18.39}\\
\hline
\multirow{6}{*}{128} & \multirow{3}{*}{NDC} & KM & 128 & 128 & 128 & 128 & 128 & 128 & 128 & 128 & 128\\
& & WSM & 7.70 & 9.11 & 9.33 & 7.47 & 9.30 & 9.90 & 10.58 & 11.17 & 9.32\\
& & KM:WSM & 16.62 & 14.05 & 13.71 & 17.13 & 13.76 & 12.93 & 12.10 & 11.45 & \textbf{13.97}\\
\cline{2-12}
& \multirow{3}{*}{Time} & KM & 2992 & 3039 & 5153 & 3096 & 17919 & 3098 & 695 & 2128 & 4765\\
& & WSM & 131 & 280 & 257 & 274 & 431 & 221 & 76 & 61 & 216\\
& & KM:WSM & 22.76 & 10.85 & 20.00 & 11.30 & 41.49 & 13.96 & 9.10 & 34.38 & \textbf{20.48}\\
\hline
\multirow{6}{*}{256} & \multirow{3}{*}{NDC} & KM & 256 & 256 & 256 & 256 & 256 & 256 & 256 & 256 & 256\\
& & WSM & 13.80 & 15.53 & 15.62 & 13.41 & 15.64 & 16.41 & 18.18 & 20.51 & 16.14\\
& & KM:WSM & 18.56 & 16.48 & 16.39 & 19.08 & 16.37 & 15.60 & 14.08 & 12.48 & \textbf{16.13}\\
\cline{2-12}
& \multirow{3}{*}{Time} & KM & 5849 & 5820 & 10048 & 5793 & 34786 & 5853 & 1347 & 4160 & 9207\\
& & WSM & 304 & 662 & 463 & 434 & 688 & 429 & 205 & 246 & 429\\
& & KM:WSM & 19.21 & 8.78 & 21.66 & 13.34 & 50.53 & 13.62 & 6.57 & 16.89 & \textbf{18.83}\\
\hline
\end{tabular}
}
\end{table}

\begin{table}[ht]
\linespread{1}
\centering
\scriptsize
{
\caption{ \label{tab_mse_32} MSE comparison of the quantization methods ($K=32$)}
\begin{tabular}{|c|c|c|c|c|c|c|c|c|}
\hline
Method & AIR & BBN & BTS & LEN & PAR & PEP & FSH & PLB\\
\hline
\hline
MC & 123.9 & 546.0 & 200.2 & 205.7 & 401.2 & 332.7 & 275.9 & 136.3\\
OTT & 92.4 & 365.8 & 156.3 & 141.4 & 252.2 & 246.5 & 157.6 & 67.7\\
OCT & 101.6 & 460.3 & 174.5 & 186.2 & 343.5 & 306.8 & 218.2 & 130.4\\
WAN & 116.9 & 509.1 & 198.5 & 216.2 & 364.7 & 333.3 & 310.5 & 111.7\\
WU & 75.3 & 421.8 & 154.5 & 157.8 & 291.1 & 264.2 & 186.6 & 68.3\\
BS & 73.6 & 388.8 & 136.5 & 138.3 & 298.9 & 261.7 & 161.4 & 89.0\\
NEU & 101.5 & 363.1 & 147.3 & 135.1 & 306.0 & 248.7 & 172.8 & 103.5\\
MMM & 134.3 & 488.9 & 203.1 & 184.9 & 331.6 & 291.8 & 234.8 & 165.6\\
SAM & 119.8 & 395.9 & 161.4 & 158.3 & 275.8 & 267.9 & 198.3 & 90.5\\
PIM & 73.9 & 412.0 & 161.4 & 159.0 & 295.5 & 265.5 & 170.0 & 134.8\\
ADU & 84.5 & 331.9 & 120.4 & 120.6 & 238.8 & 222.5 & 149.6 & 131.9\\
WSM-FGY & 58.6 & 331.3 & 117.9 & 121.0 & 235.8 & 221.7 & 147.9 & 64.1\\
WSM-LBG & 59.1 & 327.3 & 117.9 & 119.6 & 229.3 & 220.7 & 142.2 & 69.3\\
WSM-MMX & 68.5 & 331.1 & 117.9 & 126.6 & 236.7 & 225.5 & 152.6 & 103.1\\
WSM-DEN & 58.7 & 332.6 & 117.8 & 120.9 & 237.1 & 221.9 & 147.6 & 61.2\\
WSM-VAR & 59.1 & 334.9 & 116.9 & 119.1 & 229.5 & \textbf{220.1} & 147.1 & 68.5\\
WSM-SFF & 59.7 & 329.6 & 117.7 & 119.9 & 235.1 & 220.8 & 145.4 & 89.7\\
WSM-KPP & 58.9 & 330.6 & 116.5 & 120.0 & 233.1 & 221.0 & 142.3 & 74.1\\
WSM-MC & 65.0 & 335.9 & 117.7 & 119.8 & 231.4 & 221.8 & 148.1 & 71.0\\
WSM-OTT & 66.9 & 337.1 & 121.4 & 120.9 & 227.2 & 224.4 & 141.7 & 53.6\\
WSM-OCT & 58.2 & 327.1 & 117.4 & 120.9 & 239.2 & 226.7 & \textbf{141.1} & 73.4\\
WSM-WAN & 58.2 & \textbf{326.6} & 116.4 & 118.5 & 233.7 & 227.3 & 143.8 & 50.5\\
WSM-WU & \textbf{56.0} & 329.6 & \textbf{115.0} & \textbf{118.2} & \textbf{222.5} & 220.4 & 141.3 & \textbf{50.3}\\
WSM-BS & 58.9 & 332.3 & 116.9 & 118.4 & 228.0 & 221.7 & 141.9 & 54.4\\
WSM-SAM & 58.5 & 327.0 & 116.4 & 118.8 & 227.9 & 220.7 & 143.1 & 67.2\\
\hline
\end{tabular}
}
\end{table}

\begin{table}[ht]
\linespread{1}
\centering
\scriptsize
{
\caption{ \label{tab_mse_64} MSE comparison of the quantization methods ($K=64$)}
\begin{tabular}{|c|c|c|c|c|c|c|c|c|}
\hline
Method & AIR & BBN & BTS & LEN & PAR & PEP & FSH & PLB\\
\hline
\hline
MC & 81.2 & 371.0 & 126.4 & 139.2 & 258.0 & 212.8 & 169.5 & 63.7\\
OTT & 56.9 & 222.9 & 86.3 & 89.9 & 144.6 & 152.3 & 97.7 & 29.0\\
OCT & 54.4 & 269.5 & 99.7 & 109.7 & 188.2 & 179.8 & 124.7 & 48.0\\
WAN & 69.5 & 326.4 & 116.6 & 140.4 & 225.3 & 215.1 & 208.3 & 59.4\\
WU & 47.0 & 247.6 & 86.9 & 98.9 & 170.8 & 160.2 & 111.4 & 31.4\\
BS & 42.1 & 235.4 & 77.1 & 82.5 & 162.2 & 160.0 & 100.5 & 33.5\\
NEU & 46.9 & 216.2 & 79.2 & 83.4 & 153.0 & 151.1 & 107.2 & 43.8\\
MMM & 81.5 & 269.8 & 113.9 & 115.3 & 200.4 & 181.6 & 136.5 & 91.3\\
SAM & 65.4 & 245.1 & 95.3 & 102.5 & 160.5 & 160.6 & 120.1 & 54.3\\
PIM & 44.3 & 261.2 & 100.6 & 99.1 & 173.9 & 176.0 & 111.3 & 56.5\\
ADU & 43.9 & 197.6 & 66.0 & 72.9 & 132.5 & 133.4 & 90.0 & 64.0\\
WSM-FGY & 34.6 & 198.2 & 65.0 & 73.7 & 129.4 & 134.2 & 88.8 & 29.8\\
WSM-LBG & 34.0 & 196.9 & 64.0 & \textbf{71.9} & 127.6 & 131.4 & 84.6 & 28.9\\
WSM-MMX & 38.8 & 198.6 & 66.4 & 74.9 & 131.3 & 134.8 & 94.0 & 59.4\\
WSM-DEN & 34.6 & 198.3 & 65.2 & 73.8 & 129.7 & 133.7 & 89.7 & 28.4\\
WSM-VAR & \textbf{33.8} & 199.2 & 64.7 & 72.8 & 138.6 & 133.6 & 87.3 & 24.0\\
WSM-SFF & 36.8 & 198.2 & 66.3 & 72.9 & 129.9 & 133.7 & 89.9 & 46.2\\
WSM-KPP & 35.0 & 197.7 & 64.7 & 73.0 & 128.7 & 133.5 & 86.1 & 29.6\\
WSM-MC & 38.7 & 200.0 & 64.9 & 75.4 & 127.0 & 134.7 & 90.1 & 31.1\\
WSM-OTT & 41.9 & 197.3 & 67.6 & 72.4 & 127.1 & \textbf{131.3} & 86.3 & 23.7\\
WSM-OCT & 36.4 & 196.3 & 65.0 & 73.4 & 127.9 & 132.6 & 86.1 & 30.2\\
WSM-WAN & 34.2 & 197.7 & \textbf{63.4} & 71.9 & 126.0 & 134.3 & 85.0 & 22.0\\
WSM-WU & 34.3 & 196.5 & 63.5 & 72.0 & 125.2 & 131.4 & 84.7 & \textbf{21.8}\\
WSM-BS & 34.6 & 196.3 & 63.9 & \textbf{71.9} & 126.3 & 131.9 & \textbf{84.4} & 22.5\\
WSM-SAM & 35.9 & \textbf{195.4} & 64.2 & 72.9 & \textbf{125.0} & 131.5 & 86.0 & 28.3\\
\hline
\end{tabular}
}
\end{table}

\begin{table}[ht]
\linespread{1}
\centering
\scriptsize
{
\caption{ \label{tab_mse_128} MSE comparison of the quantization methods ($K=128$)}
\begin{tabular}{|c|c|c|c|c|c|c|c|c|}
\hline
Method & AIR & BBN & BTS & LEN & PAR & PEP & FSH & PLB\\
\hline
\hline
MC & 54.3 & 247.7 & 78.4 & 95.7 & 143.8 & 147.3 & 106.7 & 38.5\\
OTT & 35.9 & 144.3 & 51.8 & 57.1 & 89.5 & 97.4 & 62.3 & 14.0\\
OCT & 32.7 & 173.4 & 56.8 & 65.8 & 109.3 & 110.9 & 77.7 & 20.4\\
WAN & 50.4 & 215.6 & 70.8 & 87.4 & 146.0 & 142.1 & 124.2 & 44.9\\
WU & 30.2 & 155.3 & 50.3 & 61.5 & 96.4 & 101.0 & 68.9 & 17.4\\
BS & 26.0 & 148.2 & 43.8 & 53.1 & 95.7 & 99.7 & 64.2 & 14.2\\
NEU & 23.9 & 127.5 & 41.1 & 47.8 & 83.8 & 82.9 & 57.4 & 18.4\\
MMM & 44.0 & 188.7 & 69.1 & 73.5 & 116.9 & 113.4 & 81.0 & 42.5\\
SAM & 43.0 & 152.5 & 59.1 & 65.7 & 94.5 & 99.7 & 74.4 & 37.3\\
PIM & 29.3 & 170.8 & 63.9 & 67.5 & 107.5 & 119.4 & 79.5 & 24.4\\
ADU & 26.5 & 123.6 & 39.0 & \textbf{46.3} & 75.9 & 81.0 & 55.6 & 34.5\\
WSM-FGY & 22.2 & 125.5 & 38.2 & 47.2 & 74.6 & 81.9 & 55.8 & 13.3\\
WSM-LBG & 21.8 & 124.0 & 38.0 & 46.6 & 76.0 & 81.0 & 53.0 & 16.1\\
WSM-MMX & 24.9 & 125.5 & 39.8 & 47.9 & 77.7 & 83.2 & 59.6 & 28.9\\
WSM-DEN & 22.1 & 125.1 & 38.5 & 47.3 & 74.5 & 81.8 & 55.6 & 13.2\\
WSM-VAR & 21.9 & 124.3 & 38.1 & 47.8 & 73.2 & 81.2 & 55.3 & 12.3\\
WSM-SFF & 23.9 & 125.0 & 39.7 & 46.9 & 76.6 & 82.2 & 56.6 & 20.2\\
WSM-KPP & 22.3 & 124.5 & 38.3 & 46.6 & 75.0 & 81.2 & 53.5 & 14.3\\
WSM-MC & 24.5 & 126.4 & 38.8 & 47.9 & 76.3 & 82.7 & 60.5 & 19.1\\
WSM-OTT & 24.8 & 125.0 & 41.2 & 46.7 & 73.1 & 81.5 & 52.7 & 11.7\\
WSM-OCT & 22.9 & 124.6 & 38.9 & 46.4 & 73.7 & 81.0 & 53.7 & 12.6\\
WSM-WAN & \textbf{21.7} & \textbf{123.5} & 37.7 & 47.0 & 73.4 & 80.5 & 55.0 & 10.7\\
WSM-WU & \textbf{21.7} & 123.9 & 37.9 & 46.5 & \textbf{72.1} & 80.8 & \textbf{52.1} & 10.9\\
WSM-BS & 22.0 & 123.8 & \textbf{37.5} & \textbf{46.3} & 72.6 & \textbf{80.4} & 52.8 & \textbf{10.3}\\
WSM-SAM & 23.0 & 123.9 & 38.4 & 46.4 & 72.4 & 80.6 & 54.0 & 14.9\\
\hline
\end{tabular}
}
\end{table}

\begin{table}[ht]
\linespread{1}
\centering
\scriptsize
{
\caption{ \label{tab_mse_256} MSE comparison of the quantization methods ($K=256$)}
\begin{tabular}{|c|c|c|c|c|c|c|c|c|}
\hline
Method & AIR & BBN & BTS & LEN & PAR & PEP & FSH & PLB\\
\hline
\hline
MC & 41.1 & 165.8 & 57.2 & 65.7 & 99.5 & 98.4 & 67.7 & 26.5\\
OTT & 21.8 & 94.9 & 33.6 & 39.2 & 57.4 & 64.2 & 39.8 & 8.0\\
OCT & 19.9 & 103.8 & 32.3 & 40.5 & 63.9 & 67.3 & 44.3 & 9.2\\
WAN & 39.3 & 141.6 & 44.7 & 56.5 & 90.0 & 92.8 & 76.8 & 37.8\\
WU & 20.8 & 99.4 & 31.6 & 39.3 & 58.8 & 63.4 & 43.8 & 10.9\\
BS & 17.6 & 95.7 & 27.1 & 35.4 & 55.6 & 62.5 & 40.6 & 7.3\\
NEU & 15.5 & 84.0 & 26.4 & 31.9 & 47.5 & 54.9 & 42.2 & 9.1\\
MMM & 28.3 & 120.0 & 41.2 & 48.4 & 73.0 & 76.2 & 52.7 & 20.2\\
SAM & 31.3 & 98.6 & 41.6 & 46.0 & 60.2 & 63.9 & 48.9 & 20.2\\
PIM & 20.6 & 116.2 & 42.6 & 48.8 & 69.0 & 83.6 & 59.8 & 13.7\\
ADU & 17.7 & \textbf{78.5} & 24.3 & \textbf{30.1} & 44.6 & \textbf{50.3} & 34.9 & 20.6\\
WSM-FGY & 14.5 & 80.4 & 24.0 & 31.1 & 43.5 & 51.4 & 35.1 & 6.9\\
WSM-LBG & 14.5 & 79.5 & 23.6 & 30.5 & 44.6 & 50.8 & 33.1 & 7.3\\
WSM-MMX & 17.3 & 81.6 & 25.7 & 31.8 & 46.6 & 53.5 & 37.7 & 12.6\\
WSM-DEN & 14.6 & 80.4 & 24.0 & 31.2 & 43.6 & 51.3 & 35.3 & 7.0\\
WSM-VAR & 14.4 & 80.7 & 23.8 & 31.1 & 43.9 & 50.9 & 34.8 & 6.6\\
WSM-SFF & 16.1 & 80.8 & 25.3 & 31.1 & 45.6 & 52.3 & 36.2 & 11.1\\
WSM-KPP & 14.7 & 79.7 & 23.8 & 30.6 & 44.1 & 51.0 & 33.5 & 7.3\\
WSM-MC & 17.4 & 81.7 & 25.5 & 31.6 & 46.2 & 53.7 & 36.1 & 12.8\\
WSM-OTT & 15.3 & 80.0 & 26.6 & 31.1 & 43.5 & 51.9 & 33.7 & 6.2\\
WSM-OCT & 14.9 & 79.4 & 23.7 & 30.7 & 43.7 & 51.0 & 33.6 & 5.8\\
WSM-WAN & 14.6 & 79.4 & 23.7 & 30.5 & 43.6 & 51.1 & 34.2 & 5.9\\
WSM-WU & \textbf{14.2} & 79.0 & 23.5 & 30.4 & \textbf{42.5} & 50.4 & \textbf{32.9} & 5.9\\
WSM-BS & 14.4 & 79.2 & \textbf{23.3} & 30.3 & 42.7 & 50.6 & \textbf{32.9} & \textbf{5.6}\\
WSM-SAM & 16.2 & 79.3 & 24.6 & 31.0 & 42.8 & 50.8 & 34.1 & 7.6\\
\hline
\end{tabular}
}
\end{table}

\begin{table}[ht]
\linespread{1}
\centering
\scriptsize
{
\caption{ \label{tab_time_32} CPU time comparison of the quantization methods ($K=32$)}
\begin{tabular}{|c|c|c|c|c|c|c|c|c|}
\hline
Method & AIR & BBN & BTS & LEN & PAR & PEP & FSH & PLB\\
\hline
\hline
MC & \textbf{1} & \textbf{2} & \textbf{4} & 2 & \textbf{24} & \textbf{2} & \textbf{0} & \textbf{1}\\
OTT & 22 & 26 & 43 & 22 & 163 & 22 & \textbf{0} & 13\\
OCT & 59 & 77 & 93 & 56 & 358 & 66 & 14 & 35\\
WAN & 5 & 5 & 8 & 4 & 36 & 5 & 1 & 6\\
WU & 5 & 4 & 10 & 5 & 37 & 5 & 1 & 7\\
BS & 124 & 149 & 234 & 131 & 760 & 131 & 31 & 80\\
NEU & 86 & 74 & 128 & 70 & 466 & 72 & 11 & 47\\
MMM & 91 & 93 & 152 & 88 & 535 & 89 & 12 & 54\\
SAM & 2 & 5 & 6 & \textbf{1} & 39 & 4 & \textbf{0} & 5\\
PIM & 1406 & 1437 & 2445 & 1494 & 8687 & 1431 & 325 & 945\\
ADU & 37 & 48 & 47 & 40 & 119 & 39 & 21 & 30\\
WSM-FGY & 51 & 84 & 98 & 89 & 231 & 75 & 11 & 5\\
WSM-LBG & 74 & 160 & 203 & 154 & 477 & 123 & 22 & 14\\
WSM-MMX & 51 & 89 & 102 & 84 & 256 & 86 & 13 & 7\\
WSM-DEN & 47 & 86 & 109 & 89 & 231 & 76 & 10 & 8\\
WSM-VAR & 40 & 92 & 76 & 82 & 251 & 73 & 11 & 8\\
WSM-SFF & 46 & 75 & 88 & 68 & 222 & 68 & 11 & 4\\
WSM-KPP & 50 & 97 & 102 & 93 & 256 & 83 & 11 & 7\\
WSM-MC & 45 & 66 & 91 & 69 & 204 & 69 & 13 & 11\\
WSM-OTT & 87 & 73 & 101 & 72 & 197 & 54 & 2 & 9\\
WSM-OCT & 50 & 103 & 109 & 96 & 279 & 69 & 8 & 13\\
WSM-WAN & 35 & 80 & 73 & 56 & 214 & 50 & 7 & 5\\
WSM-WU & 29 & 66 & 84 & 70 & 204 & 49 & 6 & 8\\
WSM-BS & 59 & 98 & 110 & 84 & 339 & 87 & 20 & 31\\
WSM-SAM & 39 & 55 & 74 & 58 & 164 & 61 & 2 & 5\\
\hline
\end{tabular}
}
\end{table}

\begin{table}[ht]
\linespread{1}
\centering
\scriptsize
{
\caption{ \label{tab_time_64} CPU time comparison of the quantization methods ($K=64$)}
\begin{tabular}{|c|c|c|c|c|c|c|c|c|}
\hline
Method & AIR & BBN & BTS & LEN & PAR & PEP & FSH & PLB\\
\hline
\hline
MC & \textbf{2} & \textbf{2} & \textbf{4} & \textbf{1} & \textbf{24} & \textbf{1} & \textbf{0} & \textbf{1}\\
OTT & 32 & 38 & 52 & 30 & 194 & 31 & \textbf{0} & 19\\
OCT & 60 & 78 & 95 & 65 & 355 & 64 & 15 & 36\\
WAN & 7 & 7 & 10 & 6 & 36 & 7 & 1 & 8\\
WU & 4 & 5 & 9 & 5 & 37 & 4 & 1 & 8\\
BS & 174 & 163 & 263 & 162 & 904 & 151 & 37 & 89\\
NEU & 136 & 132 & 228 & 129 & 816 & 125 & 23 & 98\\
MMM & 151 & 167 & 286 & 153 & 943 & 148 & 24 & 125\\
SAM & 4 & 9 & 9 & 3 & 43 & 4 & \textbf{0} & 4\\
PIM & 3002 & 3018 & 5092 & 3024 & 17571 & 2909 & 688 & 2066\\
ADU & 140 & 145 & 159 & 148 & 247 & 135 & 117 & 123\\
WSM-FGY & 64 & 114 & 123 & 113 & 289 & 106 & 21 & 17\\
WSM-LBG & 125 & 233 & 256 & 239 & 636 & 218 & 40 & 28\\
WSM-MMX & 87 & 129 & 158 & 127 & 362 & 113 & 20 & 15\\
WSM-DEN & 67 & 115 & 126 & 109 & 294 & 104 & 20 & 16\\
WSM-VAR & 71 & 121 & 114 & 107 & 288 & 107 & 20 & 14\\
WSM-SFF & 70 & 101 & 120 & 97 & 291 & 90 & 17 & 17\\
WSM-KPP & 73 & 141 & 152 & 129 & 349 & 117 & 22 & 14\\
WSM-MC & 78 & 106 & 159 & 95 & 323 & 126 & 22 & 14\\
WSM-OTT & 118 & 100 & 134 & 98 & 267 & 97 & 10 & 12\\
WSM-OCT & 70 & 117 & 111 & 97 & 304 & 104 & 21 & 24\\
WSM-WAN & 49 & 99 & 96 & 92 & 264 & 78 & 18 & 14\\
WSM-WU & 49 & 88 & 106 & 89 & 246 & 79 & 18 & 14\\
WSM-BS & 77 & 130 & 148 & 118 & 434 & 116 & 29 & 40\\
WSM-SAM & 45 & 82 & 102 & 83 & 239 & 69 & 10 & 12\\
\hline
\end{tabular}
}
\end{table}

\begin{table}[ht]
\linespread{1}
\centering
\scriptsize
{
\caption{ \label{tab_time_128} CPU time comparison of the quantization methods ($K=128$)}
\begin{tabular}{|c|c|c|c|c|c|c|c|c|}
\hline
Method & AIR & BBN & BTS & LEN & PAR & PEP & FSH & PLB\\
\hline
\hline
MC & \textbf{1} & \textbf{1} & \textbf{4} & \textbf{1} & \textbf{25} & \textbf{2} & \textbf{0} & \textbf{2}\\
OTT & 44 & 47 & 68 & 41 & 246 & 43 & 1 & 23\\
OCT & 66 & 85 & 105 & 69 & 363 & 78 & 16 & 37\\
WAN & 6 & 6 & 9 & 5 & 36 & 7 & 1 & 8\\
WU & 4 & 5 & 11 & 4 & 37 & 4 & 1 & 7\\
BS & 206 & 207 & 388 & 214 & 1050 & 195 & 46 & 115\\
NEU & 253 & 247 & 420 & 241 & 1420 & 250 & 48 & 176\\
MMM & 290 & 311 & 524 & 296 & 1766 & 310 & 52 & 195\\
SAM & 8 & 32 & 12 & 8 & 57 & 14 & 2 & 8\\
PIM & 6240 & 6159 & 10429 & 6269 & 36750 & 6251 & 1373 & 4023\\
ADU & 588 & 619 & 613 & 591 & 896 & 608 & 552 & 582\\
WSM-FGY & 118 & 172 & 188 & 172 & 408 & 161 & 52 & 47\\
WSM-LBG & 228 & 407 & 425 & 378 & 970 & 357 & 104 & 64\\
WSM-MMX & 185 & 224 & 293 & 213 & 549 & 204 & 62 & 37\\
WSM-DEN & 114 & 180 & 193 & 178 & 413 & 162 & 53 & 45\\
WSM-VAR & 107 & 165 & 188 & 180 & 402 & 163 & 49 & 45\\
WSM-SFF & 121 & 167 & 226 & 154 & 425 & 158 & 50 & 40\\
WSM-KPP & 132 & 231 & 248 & 216 & 534 & 195 & 55 & 39\\
WSM-MC & 184 & 175 & 273 & 185 & 458 & 181 & 51 & 42\\
WSM-OTT & 97 & 157 & 287 & 170 & 392 & 141 & 50 & 30\\
WSM-OCT & 113 & 175 & 224 & 164 & 425 & 157 & 46 & 42\\
WSM-WAN & 93 & 149 & 197 & 136 & 402 & 150 & 41 & 39\\
WSM-WU & 92 & 152 & 139 & 149 & 404 & 144 & 42 & 30\\
WSM-BS & 125 & 192 & 212 & 175 & 588 & 180 & 53 & 63\\
WSM-SAM & 102 & 153 & 220 & 140 & 353 & 131 & 42 & 34\\
\hline
\end{tabular}
}
\end{table}

\begin{table}[ht]
\linespread{1}
\centering
\scriptsize
{
\caption{ \label{tab_time_256} CPU time comparison of the quantization methods ($K=256$)}
\begin{tabular}{|c|c|c|c|c|c|c|c|c|}
\hline
Method & AIR & BBN & BTS & LEN & PAR & PEP & FSH & PLB\\
\hline
\hline
MC & \textbf{2} & \textbf{1} & \textbf{4} & \textbf{1} & \textbf{27} & \textbf{3} & \textbf{0} & \textbf{2}\\
OTT & 53 & 62 & 86 & 51 & 326 & 56 & 10 & 25\\
OCT & 71 & 87 & 111 & 76 & 439 & 76 & 21 & 36\\
WAN & 8 & 8 & 9 & 7 & 40 & 7 & 1 & 8\\
WU & 4 & 4 & 8 & 4 & 38 & 5 & 1 & 5\\
BS & 228 & 253 & 398 & 233 & 1271 & 221 & 71 & 165\\
NEU & 471 & 462 & 786 & 431 & 2769 & 442 & 117 & 310\\
MMM & 540 & 653 & 987 & 551 & 3541 & 544 & 100 & 366\\
SAM & 11 & 67 & 15 & 9 & 80 & 30 & 5 & 14\\
PIM & 12517 & 12265 & 20951 & 12344 & 73157 & 12047 & 2775 & 7681\\
ADU & 2821 & 2900 & 2843 & 2887 & 3170 & 2791 & 2676 & 2807\\
WSM-FGY & 354 & 472 & 494 & 448 & 773 & 426 & 228 & 183\\
WSM-LBG & 559 & 806 & 942 & 770 & 1729 & 797 & 378 & 348\\
WSM-MMX & 539 & 652 & 863 & 555 & 1183 & 570 & 277 & 180\\
WSM-DEN & 345 & 464 & 490 & 445 & 828 & 422 & 232 & 199\\
WSM-VAR & 333 & 406 & 421 & 491 & 800 & 384 & 235 & 222\\
WSM-SFF & 390 & 475 & 608 & 414 & 917 & 460 & 255 & 166\\
WSM-KPP & 362 & 535 & 582 & 502 & 1018 & 478 & 229 & 168\\
WSM-MC & 419 & 453 & 773 & 498 & 1002 & 516 & 205 & 125\\
WSM-OTT & 335 & 446 & 411 & 389 & 872 & 424 & 215 & 214\\
WSM-OCT & 308 & 449 & 458 & 435 & 795 & 351 & 209 & 146\\
WSM-WAN & 335 & 449 & 442 & 460 & 746 & 356 & 241 & 142\\
WSM-WU & 281 & 371 & 362 & 335 & 654 & 324 & 196 & 132\\
WSM-BS & 302 & 427 & 471 & 421 & 891 & 409 & 229 & 217\\
WSM-SAM & 320 & 405 & 686 & 403 & 628 & 373 & 256 & 152\\
\hline
\end{tabular}
}
\end{table}

\begin{table}[ht]
\linespread{1}
\centering
\scriptsize
{
\caption{ \label{tab_mse_time_rank} MSE and CPU time rank comparison of the quantization methods}
\begin{tabular}{|c|c|c|c|c|c|c|c|c|c|c||c|}
\hline
 \multirow{2}{*}{Method} & \multicolumn{5}{|c|}{MSE Rank} & \multicolumn{5}{|c||}{CPU Time Rank} & Overall\\
 & 32 & 64 & 128 & 256 & Mean & 32 & 64 & 128 & 256 & Mean & Rank\\
\hline
\hline
MC & 24.25 & 24.25 & 24.50 & 24.75 & 24.44 & \textbf{1.13} & \textbf{1.00} & \textbf{1.00} & \textbf{1.00} & \textbf{1.03} & 12.73\\
OTT & 16.13 & 16.50 & 16.88 & 17.88 & 16.85 & 6.25 & 6.13 & 4.63 & 4.88 & 5.47 & 11.16\\
OCT & 22.13 & 21.13 & 20.63 & 19.38 & 20.81 & 15.50 & 9.75 & 6.63 & 6.00 & 9.47 & 15.14\\
WAN & 23.50 & 23.88 & 24.38 & 24.25 & 24.00 & 3.63 & 3.75 & 2.63 & 3.00 & 3.25 & 13.63\\
WU & 17.63 & 18.75 & 18.63 & 18.63 & 18.41 & 4.25 & 2.75 & 2.50 & 2.00 & 2.88 & 10.64\\
BS & 17.00 & 16.75 & 16.63 & 16.13 & 16.63 & 23.75 & 22.63 & 18.50 & 9.38 & 18.56 & 17.59\\
NEU & 18.13 & 16.88 & 14.50 & 15.25 & 16.19 & 17.25 & 20.25 & 20.38 & 17.25 & 18.78 & 17.48\\
MMM & 23.38 & 23.50 & 23.00 & 22.13 & 23.00 & 20.88 & 22.75 & 22.00 & 20.50 & 21.53 & 22.27\\
SAM & 19.88 & 20.00 & 20.13 & 20.75 & 20.19 & 2.50 & 2.63 & 4.13 & 4.13 & 3.34 & 11.77\\
PIM & 19.63 & 20.13 & 21.00 & 21.88 & 20.66 & 25.00 & 25.00 & 25.00 & 25.00 & 25.00 & 22.83\\
ADU & 14.13 & 12.50 & 10.00 & 9.25 & 11.47 & 9.75 & 19.88 & 23.50 & 23.88 & 19.25 & 15.36\\
WSM-FGY & 9.50 & 10.25 & 9.75 & 8.63 & 9.53 & 14.25 & 13.63 & 13.63 & 15.13 & 14.16 & 11.84\\
WSM-LBG & 7.00 & 4.13 & 6.63 & 5.88 & 5.91 & 22.13 & 22.00 & 22.00 & 22.50 & 22.16 & 14.03\\
WSM-MMX & 13.50 & 14.88 & 15.50 & 14.88 & 14.69 & 16.38 & 16.38 & 18.00 & 20.38 & 17.78 & 16.24\\
WSM-DEN & 10.00 & 10.00 & 9.50 & 9.13 & 9.66 & 15.13 & 13.25 & 14.63 & 15.25 & 14.56 & 12.11\\
WSM-VAR & 7.38 & 8.25 & 7.13 & 7.63 & 7.60 & 13.38 & 13.00 & 12.63 & 14.00 & 13.25 & 10.42\\
WSM-SFF & 9.13 & 11.75 & 12.50 & 13.25 & 11.66 & 10.38 & 11.50 & 13.13 & 16.50 & 12.88 & 12.27\\
WSM-KPP & 7.63 & 8.25 & 8.13 & 7.75 & 7.94 & 16.50 & 16.63 & 17.75 & 17.38 & 17.06 & 12.50\\
WSM-MC & 10.25 & 12.00 & 14.13 & 14.75 & 12.78 & 12.38 & 14.88 & 16.25 & 15.88 & 14.84 & 13.81\\
WSM-OTT & 9.25 & 7.38 & 8.63 & 8.88 & 8.53 & 12.25 & 11.25 & 10.75 & 13.00 & 11.81 & 10.17\\
WSM-OCT & 8.38 & 8.13 & 7.25 & 6.00 & 7.44 & 16.75 & 13.63 & 12.75 & 11.50 & 13.66 & 10.55\\
WSM-WAN & 5.50 & 4.50 & 4.13 & 6.25 & 5.09 & 8.50 & 8.75 & 8.75 & 12.63 & 9.66 & 7.38\\
WSM-WU & \textbf{1.88} & \textbf{2.88} & 2.75 & 2.00 & \textbf{2.38} & 9.38 & 8.50 & 8.38 & 8.13 & 8.59 & \textbf{5.48}\\
WSM-BS & 5.88 & 3.38 & \textbf{2.38} & \textbf{1.88} & 3.38 & 19.75 & 18.13 & 16.63 & 13.50 & 17.00 & 10.19\\
WSM-SAM & 4.00 & 5.00 & 6.38 & 7.88 & 5.81 & 7.50 & 6.63 & 8.63 & 12.25 & 8.75 & 7.28\\
\hline
\end{tabular}
}
\end{table}

\begin{table}[ht]
\linespread{1}
\centering
\scriptsize
{
\caption{ \label{tab_iters} Iteration count comparison of the WSM variants}
\begin{tabular}{|c|c|c|c|c|c|c|c|c|}
\hline
Method & AIR & BBN & BTS & LEN & PAR & PEP & FSH & PLB\\
\hline
\hline
$K=32$ & \multicolumn{8}{|c|}{}\\
\hline
WSM-FGY & 32 & 23 & 30 & 29 & 24 & 27 & 28 & 19\\
WSM-LBG & 6 & 4 & 18 & 3 & 16 & 8 & 5 & 15\\
WSM-MMX & 34 & 21 & 29 & 22 & 26 & 24 & 31 & 20\\
WSM-DEN & 31 & 24 & 34 & 29 & 23 & 27 & 23 & 24\\
WSM-VAR & 18 & 24 & 17 & 21 & 25 & 22 & 26 & 22\\
WSM-SFF & 35 & 19 & 25 & 19 & 23 & 22 & 25 & 16\\
WSM-KPP & 23 & 20 & 26 & 21 & 22 & 21 & 20 & 15\\
WSM-MC & 29 & 13 & 29 & 19 & 17 & 25 & 30 & 31\\
WSM-OTT & 71 & 18 & 30 & 21 & 12 & 14 & 11 & 18\\
WSM-OCT & 26 & 24 & 31 & 29 & 23 & 15 & 13 & 22\\
WSM-WAN & 18 & 21 & 18 & 12 & 19 & 12 & 18 & 12\\
WSM-WU & 9 & 13 & 23 & 18 & 16 & 9 & 13 & 16\\
WSM-BS & 14 & 13 & 10 & 11 & 8 & 13 & 10 & 10\\
WSM-SAM & 22 & 8 & 19 & 14 & 7 & 18 & 12 & 14\\
\hline
\hline
$K=64$ & \multicolumn{8}{|c|}{}\\
\hline
WSM-FGY & 28 & 25 & 29 & 28 & 25 & 30 & 29 & 20\\
WSM-LBG & 6 & 3 & 11 & 6 & 19 & 15 & 6 & 10\\
WSM-MMX & 44 & 25 & 40 & 26 & 35 & 29 & 24 & 15\\
WSM-DEN & 29 & 24 & 30 & 27 & 25 & 29 & 25 & 21\\
WSM-VAR & 32 & 26 & 23 & 22 & 20 & 27 & 25 & 21\\
WSM-SFF & 37 & 21 & 32 & 22 & 27 & 24 & 22 & 16\\
WSM-KPP & 24 & 20 & 27 & 20 & 23 & 23 & 22 & 14\\
WSM-MC & 43 & 20 & 43 & 22 & 35 & 44 & 30 & 19\\
WSM-OTT & 63 & 17 & 32 & 20 & 15 & 25 & 12 & 12\\
WSM-OCT & 30 & 20 & 18 & 17 & 16 & 24 & 19 & 27\\
WSM-WAN & 15 & 18 & 17 & 19 & 17 & 16 & 19 & 15\\
WSM-WU & 13 & 12 & 21 & 17 & 13 & 16 & 20 & 9\\
WSM-BS & 12 & 16 & 14 & 13 & 14 & 17 & 13 & 12\\
WSM-SAM & 13 & 9 & 20 & 15 & 12 & 11 & 12 & 18\\
\hline
\hline
$K=128$ & \multicolumn{8}{|c|}{}\\
\hline
WSM-FGY & 26 & 23 & 28 & 26 & 26 & 28 & 23 & 19\\
WSM-LBG & 11 & 8 & 14 & 7 & 25 & 15 & 5 & 4\\
WSM-MMX & 48 & 27 & 48 & 27 & 39 & 33 & 26 & 13\\
WSM-DEN & 26 & 24 & 27 & 26 & 25 & 27 & 24 & 19\\
WSM-VAR & 22 & 19 & 25 & 24 & 21 & 27 & 20 & 16\\
WSM-SFF & 30 & 23 & 40 & 22 & 30 & 27 & 22 & 16\\
WSM-KPP & 22 & 20 & 26 & 20 & 25 & 23 & 18 & 13\\
WSM-MC & 54 & 23 & 48 & 29 & 35 & 33 & 22 & 16\\
WSM-OTT & 18 & 18 & 51 & 24 & 17 & 20 & 22 & 10\\
WSM-OCT & 21 & 19 & 32 & 20 & 19 & 22 & 19 & 16\\
WSM-WAN & 17 & 16 & 27 & 16 & 22 & 23 & 16 & 14\\
WSM-WU & 16 & 17 & 12 & 17 & 22 & 21 & 16 & 9\\
WSM-BS & 12 & 15 & 16 & 15 & 19 & 20 & 11 & 9\\
WSM-SAM & 21 & 11 & 34 & 17 & 14 & 16 & 16 & 13\\
\hline
\hline
$K=256$ & \multicolumn{8}{|c|}{}\\
\hline
WSM-FGY & 23 & 25 & 26 & 25 & 26 & 26 & 19 & 15\\
WSM-LBG & 9 & 3 & 10 & 6 & 18 & 8 & 6 & 8\\
WSM-MMX & 36 & 33 & 49 & 28 & 43 & 32 & 22 & 14\\
WSM-DEN & 23 & 24 & 26 & 24 & 26 & 25 & 19 & 17\\
WSM-VAR & 22 & 20 & 21 & 27 & 23 & 22 & 18 & 17\\
WSM-SFF & 27 & 26 & 35 & 23 & 35 & 28 & 21 & 13\\
WSM-KPP & 19 & 20 & 23 & 20 & 25 & 22 & 17 & 13\\
WSM-MC & 29 & 23 & 47 & 28 & 40 & 32 & 17 & 10\\
WSM-OTT & 21 & 22 & 19 & 20 & 28 & 23 & 18 & 14\\
WSM-OCT & 18 & 21 & 21 & 20 & 22 & 18 & 18 & 14\\
WSM-WAN & 20 & 22 & 21 & 25 & 22 & 19 & 20 & 11\\
WSM-WU & 16 & 16 & 15 & 15 & 16 & 16 & 16 & 10\\
WSM-BS & 14 & 16 & 16 & 18 & 16 & 18 & 16 & 15\\
WSM-SAM & 20 & 15 & 40 & 21 & 14 & 19 & 22 & 13\\
\hline
\end{tabular}
}
\end{table}

\begin{table}[ht]
\linespread{1}
\centering
\scriptsize
{
\caption{ \label{tab_iters_rank} Iteration count rank comparison of the WSM variants}
\begin{tabular}{|c|c|c|c|c||c|}
\hline
Method & 32 & 64 & 128 & 256 & Mean\\
\hline
\hline
WSM-FGY & 11.75 & 11.13 & 11.38 & 10.75 & 11.25\\
WSM-LBG & 2.00 & 2.00 & 2.00 & 1.38 & 1.84\\
WSM-MMX & 11.38 & 11.50 & 12.63 & 13.25 & 12.19\\
WSM-DEN & 12.13 & 11.13 & 11.00 & 10.00 & 11.06\\
WSM-VAR & 8.88 & 10.25 & 7.63 & 8.25 & 8.75\\
WSM-SFF & 8.75 & 9.88 & 10.38 & 10.88 & 9.97\\
WSM-KPP & 7.75 & 7.50 & 7.25 & 5.88 & 7.09\\
WSM-MC & 9.00 & 11.75 & 12.13 & 10.38 & 10.81\\
WSM-OTT & 7.25 & 7.00 & 6.50 & 7.38 & 7.03\\
WSM-OCT & 9.50 & 6.88 & 6.75 & 5.38 & 7.13\\
WSM-WAN & 4.88 & 5.00 & 5.25 & 6.88 & 5.50\\
WSM-WU & 4.25 & 3.63 & 3.63 & 2.25 & 3.44\\
WSM-BS & 2.25 & 3.13 & 2.50 & 3.63 & 2.88\\
WSM-SAM & 4.00 & 3.25 & 4.13 & 6.63 & 4.50\\
\hline
\end{tabular}
}
\end{table}

\begin{table}[ht]
\linespread{1}
\centering
\scriptsize
{
\caption{ \label{tab_mse_improvement} MSE improvements for the preclustering methods}
\begin{tabular}{|c|c|c|c|c||c|}
\hline
Method & 32 & 64 & 128 & 256 & Mean\\
\hline
\hline
MC & 42\% & 47\% & 49\% & 52\% & 47\%\\
OTT & 15\% & 17\% & 19\% & 21\% & 18\%\\
OCT & 34\% & 32\% & 31\% & 27\% & 31\%\\
WAN & 44\% & 49\% & 52\% & 55\% & 50\%\\
WU & 24\% & 25\% & 26\% & 27\% & 25\%\\
BS & 19\% & 19\% & 19\% & 18\% & 19\%\\
SAM & 26\% & 30\% & 32\% & 35\% & 31\%\\
\hline
\end{tabular}
}
\end{table}

\section{Conclusions}
\label{sec_conc}
In this paper, the k-means clustering algorithm was investigated from a color quantization perspective. This algorithm has been criticized in the quantization literature because of its high computational requirements and sensitivity to initialization. We first introduced a fast and exact k-means variant that utilizes data reduction, sample weighting, and accelerated nearest neighbor search. This fast k-means algorithm was then used in the implementation of several quantization methods each one featuring a different initialization scheme. Extensive experiments on a large set of classic test images demonstrated that the proposed k-means implementations outperform state-of-the-art quantization methods with respect to distortion minimization. Other advantages of the presented methods include ease of implementation, high computational speed, and the possibility of incorporating spatial information into the quantization procedure.
\par
The implementation of the k-means based quantization methods will be made publicly available as part of the Fourier image processing and analysis library, which can be downloaded from \url{http://sourceforge.net/projects/fourier-ipal}.

\section*{Acknowledgments}
This publication was made possible by a grant from the Louisiana Board of Regents (LEQSF2008-11-RD-A-12). The author is grateful to the anonymous reviewers for their insightful suggestions and constructive comments that improved the quality and presentation of this paper.

\bibliographystyle{elsarticle-num}
\bibliography{kmeans_init_quant_bib}

\end{document}